%% file: Thesis_Demirchian.tex
\documentclass[12pt,a4paper]{article}

\input{preamble}

\begin{document}
	
\thispagestyle{empty}
\def\thefootnote{\fnsymbol{footnote}}
\vspace*{0.5cm}
\begin{center}
{\large NAS RA V. Ambartsumian's Byurakan Astrophysical Observatory}\\
\vspace*{5cm}
{\bf\Large Hovhannes Demirchian}\\
\vspace{1cm}
{\bf\Large Integrable systems connected with black holes}\\
\vspace{1cm}
{\Large PhD Thesis}\\
\vspace{1cm}
{\large 01.03.02 - Astrophysics, Radioastronomy}\\
{\large for PhD in physics and mathematics}\\
\vspace{2cm}
{\qquad\qquad\qquad\qquad\qquad\qquad\qquad\qquad\large  Adviser: Doctor of Science (Physics)}\\
{\qquad\qquad\qquad\qquad\qquad\qquad\qquad\qquad\large  Armen Nersessian}\\
\vspace{3cm}
{\large Byurakan-2019}
\end{center}

\newpage

\tableofcontents

\newpage

\include{Introduction}

\include{Gravitational_memory}

\include{Integrability_of_MP_1}

\include{Integrability_of_MP_2}

\include{Klein_Gordonization}

\include{Conclusion}

\include{Acknowledgement}

\include{Bibliography}

\end{document}

%% file: preamble.tex
\usepackage{geometry}
\usepackage{bbm,yfonts}
\usepackage[nottoc]{tocbibind} 
\usepackage{xspace}	
\usepackage{mathtools}
\usepackage{graphicx}
\usepackage{ytableau}
\usepackage{pgf,pgfarrows,pgfnodes}
\usepackage{graphics}
\usepackage[colorlinks=true,
			urlcolor=blue,
			linkcolor=black]{hyperref}
\usepackage{tikz}
\usepackage{xparse}
\usepackage{subfigure}
\usepackage{scrextend}
\usepackage{amssymb}
\usepackage{cite}

\usepackage{ifthen}	

\usepackage{titlesec}					

\titleformat{\section}[display]{\normalfont\Large\bfseries}
{CHAPTER \thesection}{0em}{\MakeUppercase} 
\titlespacing*{\section}{0em}{0em}{5em}   

\titleformat{\subsection}{\normalfont\normalsize\bfseries}
{\thesubsection}{1em}{\MakeUppercase} 
\titlespacing*{\subsection}{0em}{0em}{1em}   

\titleformat{\subsubsection}{\normalfont\normalsize\bfseries}
{\thesubsubsection}{1em}{\MakeUppercase} 
\titlespacing*{\subsubsection}{0em}{3em}{2em}   

\newcommand{\chap}[1]{\textbf{Chapter #1}}
\newcommand{\sect}[1]{\textbf{Section #1}}
\newcommand{\ssect}[1]{\textbf{Subsection #1}}

\newcommand{\beq}{\begin{equation}}
\newcommand{\eeq}{\end{equation}}
\newcommand{\beqarr}{\begin{eqnarray}}
\newcommand{\eeqarr}{\end{eqnarray}}
\newcommand{\ber}{\begin{array}}
	\newcommand{\eer}{\end{array}}
\newcommand{\del}{\partial}

\newcommand{\dsty}{\displaystyle}
\newcommand{\te}{\theta}

\graphicspath{
	{images/}
	{.}
}

\DeclareMathAlphabet{\pazocal}{OMS}{zplm}{m}{n}
\newcommand{\diff}{\operatorname{d}\!}
\newcommand{\trf}{\mathrm{tr}(f)}
\newcommand{\detf}{\mathrm{det}(f)}
\newcommand{\dd}{\mathrm{d}}
\newcommand{\ex}[1]{\mbox{e}^{\,\textstyle#1}}

\def\be{\begin{equation}}
\def\ee{\end{equation}}
\def\bea{\begin{eqnarray}}
\def\eea{\end{eqnarray}}

\def\N2{$N{=}2$}
\def\diff{{\rm d}}

\newcommand{\pdiff}{\partial}
\newcommand{\casimir}{\widetilde{\mathcal{I}}}
\newcommand{\nnr}{\nonumber \\}

\newcommand{\eps}{\varepsilon}

\def\l3{\ell_3}

\newcommand{\neweq}[2]{\ifthenelse{\equal{#1}{*}}{\begin{equation*}#2\end{equation*}}
	{\ifthenelse{\equal{#1}{}}{\begin{equation}#2\end{equation}}
		{\begin{equation}\label{#1}#2\end{equation}}}}	

\newcommand{\arrmem}[1]{#1_{\mu\nu}}	
\newcommand{\hand}[1]{$\mathcal{#1}$}	

\newcommand{\ltl}{lightlike\xspace}

\newcommand{\tml}{timelike\xspace}

\newcommand{\spl}{spacelike\xspace}

\newcommand{\hs}{hypersurface\xspace}

\newcommand{\spt}{space-time\xspace}

%% file: Introduction.tex
\section{Introduction}

Einstein equations have been discovered for over a century now and found many important applications \cite{Bicak:2000} in experimental and theoretical physics.  Despite the long period they are being studied by the scientific community, there are just a few exact solutions \cite{Bicak:2000,Negi:2004eh} known so far and one class of them is called black hole (BH) metrics. These solutions were one of the main discoveries of general relativity, first of all, due to their astrophysical importance. Black holes are assumed to be a final step in star evolution \cite{hoyle1,hoyle2,Begelman:2009ty}, are believed to make an important contribution in galaxy formation processes \cite{Cattaneo, Ferrarese:2000se,Gebhardt:2000fk} and are ``blamed'' to be responsible for a great amount of high energy radiation \cite{Blandford, Narayan:2012yp} that we detect in the universe. Although black holes have not been observed directly, their indirect observations are overwhelming \cite{Rees:1997eb, Menou:1997qd, Carr} and from general considerations it is believed that they should be rotating objects with almost no electric charge. The space-time of such a black hole is best approximated by the Kerr metric \cite{Kerr}, which is a four-dimensional stationary, asymptoticly flat vacuum solution of Einstein's equations. There exists higher dimensional generalization of Kerr space-time which is called Myers-Perry black hole \cite{mp}. Just like Kerr black hole, it is also a stationary, asymptoticly flat vacuum solution of Einstein's equations which describes a spinning black hole in an arbitrary dimension.

An important special case of Kerr black hole is the so called extremal Kerr solution which has the smallest possible mass for a given angular momentum or charge. Some astrophysical black holes have been claimed to be very close to the extremity bound, e.g. Cygnus X-1 \cite{Gou:2013dna} or MCG-6-30-15 \cite{Brenneman:2006hw}, although other independent data analyses led to opposite results \cite{Fender}. If in the future, the measurements of high angular momentum will be confirmed, extremal black holes will start to represent real astrophysical interest. 

Black hole geometries are also important objects in mathematical physics. Many of them represent a background for integrable systems. Some of these integrable systems have been unknown prior to their discovery in black hole geometries. Particularly interesting is the integrability of Hamilton-Jacobi equation as it describes the geodesics of particles. Geodesics in the near horizon limit of Kerr black hole are associated with black hole accretions which might be a source of Very High Energy (VHE) gamma-ray bursts (for a review \cite{accretion}). Accretions around black holes can also be the key to the first direct observation of a black hole \cite{Falcke:1999pj} (e.g. with the Event Horizon Telescope). 

As it is known, the Killing vectors of a geometry are associated with integrals of motion of a geodesic in that metric. In the case of the near horizon metric of an extremal rotating black hole, the killing vectors obey the structural relation of $SO$(2, 1) algebra.  It has been demonstrated (e.g. \cite{Armen-Tigran2,Hakobyan:2010ia, conformal-mechanics-BH-2_3, conformal-mechanics-BH-2_1}) that the Casimir element of this $SO$(2, 1) algebra gives rise to a reduced Hamiltonian system called spherical or angular mechanics, which contains all the specific information about the near horizon geometry. By reformulating this discussion one can say that a massive particle moving in the near horizon geometry of an extremal rotating black hole possesses  dynamical conformal symmetry, i.e.  defines ``conformal mechanics" \cite{conformal-mechanics-BH-1, GNS-11,GNS-12,  non-equal-general, Demirchian:2017uvo, conformal-mechanics-BH-2_1,conformal-mechanics-BH-2_2,conformal-mechanics-BH-2_3,conformal-mechanics-BH-2_4,conformal-mechanics-BH-2_5,conformal-mechanics-BH-2_6,conformal-mechanics-BH-2_7,conformal-mechanics-BH-2_8}, whose Casimir element can be viewed as a reduced Hamiltonian, which contains all the necessary information about the whole system.

On the other hand this reduced Hamiltonian or the spherical mechanics can be thought of as a separate system. Spherical mechanics associated with near horizon extremal black hole geometries are relatively unexplored. Latest works in this direction include \cite{conformal-mechanics-BH-2_4}, where the Hamiltonian of the spherical mechanics associated with Near Horizon Extreme Myers-Perry (NHEMP) geometry has been constructed for the special case when all rotation parameters of the black hole are equal. In \cite{GNS-11,GNS-12} the integrability of this system has been proven and the integrals of motion were presented. Extremal Myers-Perry black holes with nonequal nonvanishing rotation parameters in odd dimensions have been studied in \cite{non-equal-general} where the integrability of such systems was proven and separation of variables was carried out.


As we will see, the near horizon geometry of Myers-Perry black holes contains integrable and superintegrable systems like Rosochatius and P\"oschl-Teller systems. Studies of these kind of systems is important as they appear in many topics of theoretical physics. Another approach that we have adopted here for investigating such systems is their geometrization procedure. Geometrical counterparts of classical systems have been studied extensively. They provide a new viewpoint to existing and well-known classical systems and spread some light on their underlying structure. An important approach for geometrization of classical problems is the Jacobi metric approach \cite{jacobi1,jacobi2,jacobi3}. This is a procedure for producing a geodesic from a given Hamiltonian, which has many important applications. In particular, Onge studied the curvature of the the Jacobi metric for the Newtonian $N$-body problem \cite{Onge}, which in $N = 2$ case, reduces to the Kepler’s problem of the relative motion. 

We will propose a geometrization procedure for quantum systems. We are mostly interested in problems which are superintegrable in higher dimensions. Particularly interesting are the Higgs oscillator \cite{Higgs,Leemon}, which is a particle on a $d$-sphere with a specific potential and the superintegrable Rosochatius system - a direct generalization of the Higgs oscillator. We will encounter the classical superintegrable Rosochatius system in \sect{\ref{fully-isotropic-sec}} as the angular mechanics of near horizon limit of fully isotropic Myers-Perry black hole. Separation of variables in Rosochatius system results into a recursive family of one-dimensional P\"oschl-Teller system. 

Higgs oscillator, Rosochatius system and P\"oschl-Teller system belong to a class of quantum quantum systems where energy are quadratic functions of the energy level number. After the geometrization procedure proposed in \chap{\ref{sec:klein_gordon}}, these systems will result into Klein-Gordon equations with eigenmode frequencies linear in the frequency level number. In other words this means that the frequencies are highly resonant, which itself has important consequences in the AdS stability problem (see \cite{review} for a review).


Another important class of solutions of Einstein's equations are gravitational waves. Compared to black holes, gravitational waves have been directly detected in 2015 by two LIGO (Laser Interferometer Gravitational-wave Observatory) detectors. The existence of gravitational waves has been proven by indirect astrophysical observations long before their detection. In particular, the presence of gravitational waves was confirmed by monitoring the orbital parameters of the binary millisecond pulsar PSR B1913+16 \cite{Taylor:1979zz}. Because of the gravitational radiation, the objects in the binary system lose energy and angular momentum which was detected and corresponded to the quantitative predictions of the theory. Besides being a real physical phenomena and one of the most important predictions of the theory of general relativity, gravitational waves will take an important role in observational astronomy. Compared to other types of radiation, e.g. photons, neutrinos and cosmic rays, gravitational waves don't get refracted by gas clouds and absorbed by cosmic bodies and can travel big distances, pointing directly back to the source. The importance of gravitational wave detectors will grow with their sensibility.

When a pair of inertial test particles encounter gravitational  waves, their relative positions get shifted permanently. This phenomena is called the gravitational memory effect. It is known to be related to the theory of soft gravitons and symmetries of null infinity of asymptotically flat spaces and particularly black holes. 

In \chap{\ref{sec:gr_memory}} we are going to discuss this effect and suggest its covariant formulation in frames of a model of impulsive gravitational waves. This model assumes that the space-time is divided into two domains by a hypersurface, which in general can contain a mixture of gravitational waves and other material sources. There are many examples of physical systems in nature which can be described in frames of this model. Such systems may appear after cataclysmic astrophysical events, such as a supernova or a collision of neutron stars. These systems are used to simulate an exploding white hole, to model an impulsive null signal from a system of neighboring test particles and have many other applications. In general one can choose these two metrics to be either continuous or discontinuous on the boundary \hs subject to the condition that the induced metric is unique, but either way the metric's transverse derivative will not be continuous. This always leads to a singularity in the form of a $\delta$-function in the Riemann tensor.\par
Solutions to the Einstein equations give different results depending on whether the boundary surface is taken to be  null or \tml (\spl).  In the first case, when the \hs is null, both Weyl and Ricci parts of the Riemann tensor are singular. As it is known, the Weyl part of the Riemann tensor is associated with gravitational waves, whereas the Ricci tensor has non-zero value only in the presence of some material source. Hence, in the case of null boundary \hs, both a material source and an impulsive tidal wave can be present. When the boundary surface is \tml, only the Ricci tensor is singular, giving rise only to a matter stress-energy tensor. Depending on the matter distribution, the discussed hypersurfaces are classified into two types: shock waves or boundary surfaces, which arise when there is a jump discontinuity in the density of the stress-energy tensor between the two metrics that they divide and surface layers, otherwise called thin shells, where the density becomes infinite.\par
There are two different approaches to describe singular hypersurfaces. The first one is called the distributional method. In this case a common set of coordinates is used for both sides of the \hs. The other method is a generalization of the ``cut and paste'' approach of Penrose. Here, the \spt coordinates on the two sides of the \hs can be chosen independently from each other, so in this sense it is a more general approach than the distributional algorithm. It was introduced by Israel to describe \tml hypersurfaces \cite{Israel66} but it was not suitable for the case of null hypersurfaces. In the \tml case, the Israel approach uses the extrinsic curvature of the \hs to describe the stress-energy tensor in it. When we move to the null case, the intrinsic metric of the \hs \spt becomes degenerate, because  the normal vector becomes tangent and there is no distinguishable transverse vector defined. Hence, the extrinsic curvature, which is defined in terms of the metric, is no longer uniquely definable, so  it cannot be used to study the \hs. This problem was solved and the approach was generalized  for the \ltl case by Barrab\`es and Israel \cite{Barrabes:1991ng}.


In this thesis we are going to discuss different problems related to asymptotic flat spaces, integrable systems and mathematical physics associated with black holes. First, we will study gravitational memory effect which is known to have deep connections with soft gravitons and symmetries of null infinity of asymptotically flat spaces. Then, we will discuss Myers-Perry black holes, more particularly, the near horizon geometry and associated integrable systems. Finally, we will propose a geometrization procedure for a special class of quantum (super)integrable systems, which appear in many topics of mathematical physics ( including in near horizon geometry of Myers-Perry black hole).

This thesis is based on the papers \cite{OLoughlin:2018ebk,Demirchian:2018xsk,Evnin:2017vpc,Demirchian:2017uvo, Demirchian_PAN, Demirchian_PPN}. The research done in \chap{\ref{sec:gr_memory}} was carried out under supervision of Martin O'Loughlin. The problem addressed in \chap{\ref{sec:klein_gordon}} was suggested and solved in cooperation with Oleg Evnin.

It is organized as follows.

In \chap{\ref{sec:gr_memory}} we study the effects that an impulsive signal in a singular \hs can have on a particle which encounters it. A similar question has been discussed by Barrab\`es and Hogan in \cite{Barrabes:2001vy} for the case of \tml particles, where they have constructed the geodesics deviation vector in the first order approximation and found a relation between the geodesic deviation vector, the stress-energy content and gravitational wave components of the shell. We propose a new approach for studying the effect of null shells on null geodesic congruences. This is an exact method which allows one to easily calculate the change in the expansion, shear and rotation of the congruence upon crossing the shell and its evolution to the future of the shell. We find that the effect of the shell on the congruence, as already observed in the time-like case in \cite{Barrabes_Hogan-book}, is a discontinuity in the B-tensor (the gradient of the geodesic vector). We call this the B-memory effect, which is a more covariant way of describing the gravitational memory effect. Gravitational memory effect has deep connections with soft gravitons \cite{Strominger:2014pwa}, which in turn is linked to the symmetries of null infinity of asymptotically flat spaces \cite{HPS1,HPS2}. We found the explicit relation of B-memory with the stress energy and gravitational wave components of the shell. We consider the simplest case of a null shell representing an outgoing gravitational wave and parametrized by a general soldering transformation (a subclass of which are the BMS supertranslations) in Minkowski space, but our method is applicable to any geodesic congruence that crosses a null shell localized on a killing horizon.

In \chap{\ref{sec:Integrability_of_MP_1}} and \chap{\ref{sec:Integrability_of_MP_2}} conformal mechanics associated with Near Horizon geometry of Extremal Myers-Perry (NHEMP) black hole has been studied. First, a unified description of an arbitrary odd and even dimensional geometry and conformal mechanics has been proposed. Then, the question integrability of special cases of fully non-isotropic and fully isotropic cases has been addressed in this description. We have found a non-trivial transformation from non-isotropic NHEMP conformal mechanics to its isotropic case. Furthermore, the general case, when groups of equal and non-equal rotation parameters exist, has been studied and shown that this problem reduces to its special cases of fully non-isotropic and fully isotropic NHEMP conformal mechanics. At the end of the \chap{\ref{sec:Integrability_of_MP_2}} another non-trivial near-horizon geometry has been discussed. The so-called Near Horizon Extremal Vanishing Horizon Myers-Perry black hole (NHEVHMP) is obtained when one of the rotation parameters of the Myers-Perry black hole vanishes. We studied the integrability properties of NHEVHMP in higher dimensions in fully isotropic, fully non-isotropic and general cases.

In \chap{\ref{sec:klein_gordon}} we propose a geometrization procedure which associates to a non-relativistic quantum particle in a potential on a curved spacetime a purely geodesic motion in another spacetime. In other words, we propose a correspondence between the solutions of Schroedinger equation and Klein-Gordon equation on a corresponding manifold, which itself, as it is well known, reduces to a geodesic equation through quasi-classical of Eikonal approximation. We will explain this procedure on the example of the Higgs oscillator and superintegrable Rosochatius system.

%% file: Gravitational_memory.tex
\section[Geodesic congruences, impulsive gravitational waves and gravitational 
		\\memory]
		{Geodesic congruences, 
			\\ impulsive gravitational waves
			\\ and gravitational memory}
\label{sec:gr_memory}
The study of impulsive gravitational waves in the form of null shells has recently received renewed attention due to their possible role in the transfer of information from black hole horizons to null infinity. As the black hole horizon is a killing horizon, there is an infinite variety of ways to attach (solder) the black hole interior to the black hole exterior creating a null shell on the horizon \cite{Barrabes:1991ng,Blau:2015nee}. A subclass of these can be shown to correspond to BMS like supertranslations. Furthermore the long studied BMS supertranslations at null infinity of asymptotically flat spaces are linked to the physics of soft gravitons which appear to play an important role in restoring information not seen in the hard gravitons of Hawking radiation \cite{HPS1,HPS2}. In turn the soft gravitons are related to the gravitational memory effect \cite{Strominger:2014pwa}.

Gravitational memory \cite{grav_memory,Strominger:2017zoo,Ashtekar:2018lor} is the classical change in nearby geodesics in an asymptotically flat region of space-time as they pass through an outgoing gravitational wave. The study of the effect of a null shell on a time-like congruence that crosses it has been addressed by Barrabes and Hogan \cite{Barrabes_Hogan-book, Barrabes:2001vy}. They calculated the change in the tangent vector and the geodesic deviation vector together with the expansion, shear and rotation upon crossing an impulsive gravitational wave and found a jump in the acceleration of the geodesic and derivatives of the geodesic deviation vector proportional to the stress-energy content and gravitational wave components of the shell. 

To further understand the relationship between gravitons and gravitational memory it is thus important to study the effect of waves on null geodesic congruences, not only as the congruence crosses the wave but also the future evolution of the congruence. In this chapter we describe a new exact approach for studying the effect of null shells on null geodesic congruences. This method allows one to easily calculate the change in the $\pazocal{B}$-tensor, which encodes the expansion, shear and rotation of the congruence, upon crossing the shell and its evolution to the future of the shell. We find that the effect of the shell on the congruence, as already observed in the time-like case in \cite{Barrabes_Hogan-book}, is a discontinuity in the $\pazocal{B}$-tensor, which we will refer to as $\pazocal{B}$-memory effect, not to be confused with the B-mode gravitational memory. We show how this $\pazocal{B}$-memory is determined by the stress energy and gravitational wave components of the shell. We consider the simplest case of a null shell representing an outgoing gravitational wave and parametrised by a general soldering transformation (a subclass of which are the BMS supertranslations) in Minkowski space, but our method is applicable to any geodesic congruence that crosses a null shell localised on a killing horizon. It is intriguing to note that our formulation of $\pazocal{B}$-memory has much in common with gravitational memory as formulated in \cite{Ashtekar:2018lor}.

In \sect{\ref{sec:impulsive_review}} we give a short review of the model of  impulsive signals and the distributional algorithm for constructing null singular shells. 

In \sect{\ref{subsec:memory}} we introduce the concept of a B-memory effect as a covariant formulation of gravitational memory. 

In \sect{\ref{subsec:setup}} we describe the setup of the problem and give a general description of the suggested approach. A detailed discussion of the approach is carried out in \sect{\ref{subsec:application}} while in \sect{\ref{subsec:behaviour}} the detailed behavior of a lightlike congruence is studied. 

In \sect{\ref{subsec:discussion}} we discuss our results and their relation to other formulations of gravitational memory, in particular to that reviewed in \cite{Ashtekar:2018lor}.

The results of this chapter were obtained in cooperation with Martin O'Loughlin and are based on \cite{OLoughlin:2018ebk}.

\subsection{Impulsive signals in null hypersurfaces}
\label{sec:impulsive_review}
In this section we will discuss a \spt manifold \hand{M} which is divided into two domains by a null \hs with a $C^0$ metric tensor (metric tensor is continuous across the \hs but its first derivatives are not). We will denote the domain on the left side of the \hs by \hand{M^+} and on the right side of the \hs by \hand{M^-} (see fig. \ref{cts-lines}). Let $x^{\mu}$ ($\mu$ = 0, 1, 2, 3) be the local coordinates on both sides of the \hs and $\Phi(x)=0$ be the equation of the \hs, with $\Phi>0$ corresponding to \hand{M^+} and $\Phi<0$ to \hand{M^-}. 
The normal vector to the \ltl hypersurface is 
\begin{equation}
n^\mu=\chi^{-1}(x)g^{\mu\nu}\pdiff_{\nu}\Phi(x),
\qquad
n\cdot n \equiv \arrmem{g}n^\mu n^\nu|_{\pm}=0,
\end{equation}
where $\arrmem {g^{\pm}}$ are the components of the metric tensor on \spt \hand{M^+} $\bigcup$ \hand{M^-} and $\chi$ is an arbitrary function. Any tensor field will be denoted by $+$ or $-$ superscripts on \hand{M^+} and \hand{M^-}. If these tensors differ on each side of the boundary \hs \hand{N}, the jump $[F]=F^+|_\mathcal{N}-F^-|_\mathcal{N}$ across \hand{N}, will not be zero and is an important quantity for our later derivations. Here, the subscript $\mathcal{N}$ indicates that $F^{\pm}$ should be evaluated on the hypersurface. We also define a hybrid tensor $\tilde{F}$ as follows
\begin{equation}
	\label{eq:hybrid_null}
	\tilde{F}(x)=F^+\theta(\Phi)+F^-(1-\theta(\Phi)),
	\qquad
	\theta(\Phi)=\left\{ \begin{array}{ccc}
	1\text{\quad}\Phi>0\\
	\frac{1}{2}\text{\quad}\Phi=0\\
	0\text{\quad}\Phi<0
	\end{array}\right. ,
\end{equation}
where $\theta(\Phi)$ is the Heaviside step function.
We will assume we are dealing with a continuous metric across \hand{N}, so 
\begin{equation}
	\label{eq:metric_light}
	\tilde{g}_{\mu\nu}=g_{\mu\nu}\text{\qquad and \qquad}[g_{\mu\nu}]=0.
\end{equation}
The metric is also continuous in the derivatives tangent to the \hs, but is discontinuous in the transversal derivative. We introduce symmetric tensor $\arrmem{\gamma}$ to describe the discontinuity in transversal derivative
\begin{equation}
	\label{eq:metric_deriv_jump_null}
	[\pdiff_\alpha \arrmem{g}]=\eta n_\alpha\arrmem{\gamma}.
\end{equation}
As $n_\alpha$ is a null vector, to find the components of $\arrmem{\gamma}$ we will need to introduce a transversal vector $N$, which points out of \hand{N}
\begin{equation} 
	\label{eq:transverse}
	N\cdot n\equiv\eta^{-1}\neq0, \qquad N^+_\mu=N^-_\nu\equiv N_\mu.
\end{equation}
So, it follows from \eqref{eq:metric_deriv_jump_null} that
\begin{equation}
	\label{eq:gamma_definition}
	N^\alpha[\pdiff_\alpha\arrmem{g}]=\arrmem{\gamma}.
\end{equation} 
We only restrict the choice of $\arrmem{\gamma}$ by requiring it to have a uniquely defined projection on the \hs \hand{N}. Thus, we have the following gauge freedom:
\begin{equation}
	\label{eq:gauge_null}
	\arrmem{\gamma} \rightarrow \arrmem{\gamma^\prime}=\arrmem{\gamma}+v_\mu n_\nu+n_\mu v_\nu,
\end{equation}
where $v$ is a four-dimensional vector field defined on \hand{N}. $N$ is not uniquely defined as well. The scalar product \eqref{eq:transverse} is invariant under the gauge transformation 
\begin{equation}
	\label{eq:transverse_freedom}
	N\rightarrow N^\prime=N+v,
\end{equation}
Our aim is to construct the Riemann and Einstein tensors for the \spt \hand{M}, which depend on partial derivatives of $g_{\mu\nu}$. For a partial derivative of some general $\tilde{F}$ tensor field we can write
\begin{equation}
\label{eq:hybrid_inter}
\begin{aligned}
\pdiff_\mu\tilde{F}&=\theta(\Phi)\pdiff_\mu F^{+} +F^{+}\pdiff_\mu\theta(\Phi) +(1-\theta(\Phi))\pdiff_\mu F^{-}- F^{-}\pdiff_\mu\theta(\Phi)\\
&=\tilde{\pdiff_\mu F}+[F]\pdiff_\mu \theta.
\end{aligned}
\end{equation}
For the derivative of the Heaviside step function, we can write
\begin{equation*}
\pdiff_\mu \theta(\Phi)=\pdiff_\Phi\theta(\Phi)\pdiff_\mu\Phi=\delta(\Phi)\chi n_\mu,
\end{equation*}
so equation \eqref{eq:hybrid_inter} becomes
\begin{equation}
\label{eq:deriv_hybrid_tensor}
\pdiff_\mu\tilde{F}=\tilde{\pdiff_\mu F}+[F]\chi n_\mu\delta(\Phi).
\end{equation} 
In addition, we should also note that if we have two $F$ and $G$ tensors defined, then from \eqref{eq:hybrid_null} it follows that \[\tilde{F}\tilde{G}=\tilde{FG}-[F] [G]\, \theta(\Phi)\, \theta(-\Phi).\]
To derive the form of Einstein's tensor we need to start from the Christoffel symbols. Using \eqref{eq:metric_light} and \eqref{eq:deriv_hybrid_tensor} we can obtain
\begin{equation}
	\pdiff_\rho \arrmem{\tilde{g}}=\tilde{\pdiff_\rho \arrmem{g}},
\end{equation}
from which it follows that
\begin{equation}
\label{eq:christ_inter}
\begin{aligned}
	\arrmem{\Gamma^\lambda}&=\frac{1}{2}g^{\lambda\rho}[\pdiff_\mu\tilde{g}_{\rho\nu}+\pdiff_\nu\tilde{g}_{\rho\mu}-\pdiff_\rho\tilde{g}_{\mu\nu}]\\
	&=\frac{1}{2}g^{\lambda\rho}[\tilde{\pdiff_\mu g_{\rho\nu}}+\tilde{\pdiff_\nu g_{\rho\mu}}-\tilde{\pdiff_\rho g_{\mu\nu}}]\\
	&=\arrmem{\tilde{\Gamma}^\lambda}
\end{aligned}
\end{equation}
For $[\arrmem{\Gamma^\lambda}]$ we will get
\begin{equation}
\label{eq:christ_inter_1}
\begin{aligned}
	\left[\arrmem{\Gamma^\lambda}\right]&=\frac{1}{2}g^{\lambda\rho}([\pdiff_\mu{g}_{\rho\nu}]+[\pdiff_\nu{g}_{\rho\mu}]-[\pdiff_\rho{g}_{\mu\nu}])\\
	&=\frac{1}{2}g^{\lambda\rho}(\eta n_\mu \gamma_{\rho\nu}+\eta n_\nu \gamma_{\rho\mu}-\eta n_\rho \gamma_{\mu\nu})\\
	&=\eta\left(\gamma^\lambda_{(\mu} n_{\nu)}-\frac{1}{2}\arrmem{\gamma}n^\lambda\right)
\end{aligned}
\end{equation}
The brackets around indices denote symmetrization over those indices.

Now we can calculate the Riemann tensor. The Riemann tensor is expressed through Christoffel symbols according to the following equation.
\begin{equation}
	\label{eq:reimann_christ}
	R_{k\lambda\mu\nu}=\pdiff_\mu\tilde{\Gamma}_{k\lambda\nu}-\pdiff_\nu\tilde{\Gamma}_{k\lambda\mu}+
	\tilde{\Gamma}_{k\mu\rho}\tilde{\Gamma}^{\rho}_{\, \nu\lambda}-\tilde{\Gamma}_{k\nu\rho}\tilde{\Gamma}^{\, \rho}_{\mu\lambda}
\end{equation}
The tilded Riemann tensor will be
\begin{equation}
	\begin{aligned}
	\tilde{R}_{k\lambda\mu\nu}
	&=\tilde{\pdiff_\mu{\Gamma}}\hphantom{}_{k\lambda\nu}-\tilde{\pdiff_\nu{\Gamma}}\hphantom{}_{k\lambda\mu}+
	\tilde{{\Gamma}_{k\mu\rho}{\Gamma}}\hphantom{}^{\rho}_{\, \nu\lambda}-\tilde{{\Gamma}_{k\nu\rho}{\Gamma}}\hphantom{}^{\rho}_{\, \rho\mu\lambda}
	\end{aligned}
\end{equation} 
From \eqref{eq:deriv_hybrid_tensor} we can deduce that \[\tilde{\pdiff_\mu{\Gamma}}\hphantom{}_{k\lambda\nu}=\pdiff_\mu\tilde{\Gamma}_{k\lambda\nu}-[{\Gamma}_{k\lambda\nu}]\chi n_\mu\delta(\Phi).\] So, plugging this into the tilded Riemann equation we find
\begin{equation}
\label{eq:riemann_hybrid_inter}
\begin{aligned}
	\tilde{R}_{k\lambda\mu\nu}
	&= \pdiff_\mu\tilde{\Gamma}_{k\lambda\nu}
	-\pdiff_\nu\tilde{\Gamma}_{k\lambda\mu}
	+\tilde{{\Gamma}}_{k\mu\rho}\tilde{{\Gamma}}\hphantom{}_{\rho\nu\lambda}
	-\tilde{{\Gamma}}_{k\nu\rho}\tilde{{\Gamma}}\hphantom{}_{\rho\mu\lambda}\\
	&\quad-[{\Gamma}_{k\lambda\nu}]\chi n_\mu\delta(\Phi)
	+[{\Gamma}_{k\lambda\mu}]\chi n_\nu\delta(\Phi)\\
	&=R_{k\lambda\mu\nu}-[{\Gamma}_{k\lambda\nu}]\chi n_\mu\delta(\Phi)+[{\Gamma}_{k\lambda\mu}]\chi n_\nu\delta(\Phi),
\end{aligned}
\end{equation} 
where terms that vanish distributionally have been ignored.

The last two terms in tilde-Riemann equation can be simplified.
\begin{equation*}
\begin{aligned}
	\left[{\Gamma}_{k\lambda\mu}\right]\chi n_\nu\delta(\Phi)&-[{\Gamma}_{k\lambda\nu}]\chi n_\mu\delta(\Phi)\\
	&=
	\eta\chi \delta(\Phi)[\gamma_{k(\lambda} n_{\mu)}n_\nu-\frac{1}{2}\gamma_{\lambda\mu} n_kn_\nu
	-\gamma_{k(\lambda} n_{\nu)}n_\mu+\frac{1}{2}\gamma_{\lambda\nu} n_kn_\mu]\\
	&=-\eta\chi \delta(\Phi){\hat R}_{k\lambda\mu\nu},
\end{aligned}
\end{equation*}
where 
\begin{equation*}
\begin{aligned}
	{\hat R}_{k\lambda\mu\nu}&\equiv
	\gamma_{k(\lambda} n_{\mu)}n_\nu-\frac{1}{2}\gamma_{\lambda\mu} n_kn_\nu
	-\gamma_{k(\lambda} n_{\nu)}n_\mu+\frac{1}{2}\gamma_{\lambda\nu} n_kn_\mu\\	
	&=2n_{[k}\gamma_{\lambda][\mu}n_{\nu]}
\end{aligned}
\end{equation*}
and square brackets around indices denote skew-symmetrization. Plugging this result into \eqref{eq:riemann_hybrid_inter} will give
\begin{equation}
	R_{k\lambda\mu\nu}=\tilde{R}_{k\lambda\mu\nu}+{\hat R}_{k\lambda\mu\nu}\eta\chi\delta(\Phi)
\end{equation}
Similarly we find
\neweq{}{\arrmem{R}=\arrmem{\tilde{R}}+\arrmem{\hat{R}}\epsilon\chi \delta(\Phi), \qquad
\arrmem{\hat{R}}=\gamma_{(\mu}n_{\nu)}-\frac{\gamma}{2}n_\mu n_\nu,}
\begin{equation}
	\arrmem{G}=\arrmem{\tilde{G}}+\arrmem{\hat{G}}\epsilon\chi \delta(\Phi), \qquad
	\arrmem{\hat{G}}=\gamma_{(\mu}n_{\nu)}-\frac{\gamma}{2}n_\mu n_\nu -\frac{\gamma^\dagger}{2}\arrmem{g},
\end{equation}
where
\begin{equation}
\label{eq:gammas_null}
	\gamma\equiv g^{\mu\nu}\arrmem{\gamma}, \qquad \gamma_\mu\equiv\arrmem{\gamma}n^\nu, \qquad \gamma^{\dagger}\equiv\arrmem{\gamma}n^\mu n^\nu=\gamma_\mu n^\mu
\end{equation}
From the form of the Einstein tensor we conclude that the stress-energy tensor will contain two terms one of which proportional to the Dirac $\delta$-function.
\begin{equation}
\label{eq:stress_energy_null}
\arrmem{T}=\arrmem{\tilde{T}}+\arrmem{S}\eta\chi \delta(\Phi).
\end{equation}
The tilde-term of the stress-energy tensor corresponds to the matter content $\arrmem{T^\pm}$ of the exterior domains \hand{M^\pm}. The second term corresponds to the matter in the singular \hs, which is actually a shell of \ltl matter. The stress-energy tensor on the null shell is \[\arrmem{T}|_\mathcal{N}=\arrmem{S}\eta\chi\delta(\Phi)\]
where $\arrmem{S}$ is given by 
\begin{equation}
\label{eq:stress_energy_null_distributional}
16\pi\arrmem{S}=-\gamma n_\mu n_\nu-\gamma^{\dagger}\arrmem{g}+2\gamma_{(\mu}n_{\nu)}=2\arrmem{\hat{G}}.
\end{equation}
The three terms in the stress-energy tensor in the last equation, if taken on the \hs \hand{N}, can be interpreted as being related to the energy density, the isotropic tensor and the energy current respectively.

\subsection{$\pazocal{B}$-memory as gravitational memory}
\label{subsec:memory}

The gravitational memory effect is the change in relative velocity between neighboring geodesics after the passing of a gravitational wave - the idea being that the passing of a gravitational wave leaves some ``memory'' in the relative movement of inertial observers. Here we propose a more covariant characterization of this memory effect by considering the effect of an outgoing wave in the form of a null shell on a null geodesic congruence.

To see explicitly how this works we begin with the general construction and notation of \cite{Blau:2015nee}. The impulsive wave (null shell) is confined to a singular null hypersurface $\pazocal N$ which divides the space-time into two domains $\pazocal M^-$ $\bigcup$ $\pazocal M^+$ - the past and future domains - each with its own coordinate system $x^\mu_\pm$. Each domain has its own metric, $g_{\mu\nu}^-$ or $g_{\mu\nu}^+$, together with junction conditions for soldering that relate the two metrics where they meet on the hypersurface $\pazocal N$.
The soldering determines the constituents of the impulsive wave and in the case that $\pazocal N$ coincides with a killing horizon an infinite variety of solderings are allowed \cite{Blau:2015nee} producing an infinite variety of impulsive signals. For explicit calculations we will use the freedom to perform independent coordinate transformations on $\pazocal M^-$ and $\pazocal M^+$ to choose a global coordinate system $x^\mu$ that is continuous across $\pazocal N$ and such that the metric is also continuous
\be
[g_{\mu\nu}] = g_{\mu\nu}^+ - g_{\mu\nu}^- = 0.
\ee
In these global coordinates the hypersurface $\pazocal N$ is defined by the equation $\Phi(x)=0$ with $\Phi(x)>0$ covering the future domain and $\Phi(x)<0$ covering the past domain. 

We will consider a congruence with tangent vector field $T$ transverse to $\pazocal N$ together with the null generator $n$ of the shell, where $T\cdot n=-1$, and to calculate the independent components of the $\pazocal{B}$-tensor $\pazocal{B}_{\alpha\beta} = \nabla_\beta T_\alpha$ \cite{Poisson} we will project it onto the spatial submanifold of the shell defined by a pair of space-like orthonormal vectors $e_A^\alpha$, $A\in(x,y)$ such that $e_A\cdot n=e_A\cdot T = 0$. Furthermore we can and will choose $e_A^\alpha$ to be parallel transported along the congruence, a choice that simplifies the following equations by eliminating the connection from the evolution equation for $\pazocal{B}$. The projection of $\pazocal{B}_{\alpha\beta}$ onto the congruence is
\be
\pazocal{B}_{AB} = e_A^\alpha e_B^\beta \pazocal{B}_{\alpha\beta} = \frac{1}{2}\theta\delta_{AB}+ \sigma_{AB}+\omega_{AB},
\ee
where the expansion, shear and rotation are explicitly given by
\be
\label{eq:BAB}
	\theta = \pazocal{B}^A_A\qquad
	\sigma_{AB} = \pazocal{B}_{(AB)} - \frac12 \theta\delta_{AB}
	\qquad
	\omega_{AB}=\pazocal{B}_{[AB]}.
\ee

The evolution equation for $\pazocal{B}_{AB}$ (with respect to the affine parameter $\lambda$ of the congruence) is
\be
\label{eq:Bevolution}
\frac{\dd \pazocal{B}_{AB}}{\dd \lambda} = -\pazocal{B}_{AC}\pazocal{B}_B^C -R_{AB}
\ee
and
\be
R_{AB} = R_{\alpha\mu\beta\nu}e_A^\alpha T^\mu e_B^\beta T^\nu = \frac{1}{2}\mathcal{R}\delta_{AB} + C_{AB}
\ee
where
\be
\mathcal{R} = R_{\alpha\beta}T^\alpha T^\beta \qquad C_{AB} = C_{\alpha\mu\beta\nu}e_A^\alpha T^\mu e_B^\beta T^\nu
\ee
and $C_{AB}$ is traceless.

In the presence of a null shell the Riemann and Weyl tensors have a term that
is localised on the shell and proportional to a delta function \cite{Barrabes:1991ng}. Thus we separate $\mathcal{R}$ and $C_{AB}$ into their bulk and shell components
\be
\mathcal{R}=\hat{\mathcal{R}} + \bar{\mathcal{R}}\delta(\Phi)\qquad C_{AB} = \hat{C}_{AB} + \bar{C}_{AB}\delta(\Phi),
\ee
								   
In the evolution equation for $\pazocal{B}_{AB}$ the delta function in $R_{AB}$ on the right hand side can only be balanced by a delta function in the derivative of $\pazocal{B}_{AB}$ 
meaning that the $\pazocal{B}$-tensor must be discontinuous across the shell. This discontinuity is related to the stress-energy and gravitational wave components of the shell as we will see in detail in the following sections. 

The evolution of the rotation is simply given by
\be
\frac{\dd\omega_{AB}}{\dd \lambda} = -\theta \omega_{AB},
\ee
which can be integrated to give
\be
\omega_{AB} = K \ex{-\int_{\lambda_0}^\lambda \theta\dd \lambda^\prime}\epsilon_{AB}.
\ee
We can deduce from this equation that the rotation must be continuous but not necessarily differentiable across the shell as the expansion is at most discontinuous. In particular, and as we will see in detail in the following sections, a zero rotation before the shell and at worst a finite jump in the expansion will result in zero rotation after the shell. This means that a congruence that is hypersurface orthogonal to the past of the shell must also be hypersurface orthogonal to the future of the shell.

Our calculations thus indicate that an alternative and generally covariant formulation of the gravitational memory effect is that there is a discontinuity in the $\pazocal{B}$-tensor of a congruence upon crossing a null shell. In the following sections we will show how to explicitly calculate the evolution of the $\pazocal{B}$-tensor for a congruence that crosses a null shell.

\subsection{The setup and proposal}
\label{subsec:setup}
Our general construction is applicable to any null shell located on a killing horizon. For simplicity (and without loss of conceptual insight) we will consider in the following sections exclusively the case of a planar null hypersurface (which is obviously a killing horizon) in Minkowski space.

\begin{figure}[h!]
	\center{\includegraphics[width=7cm]{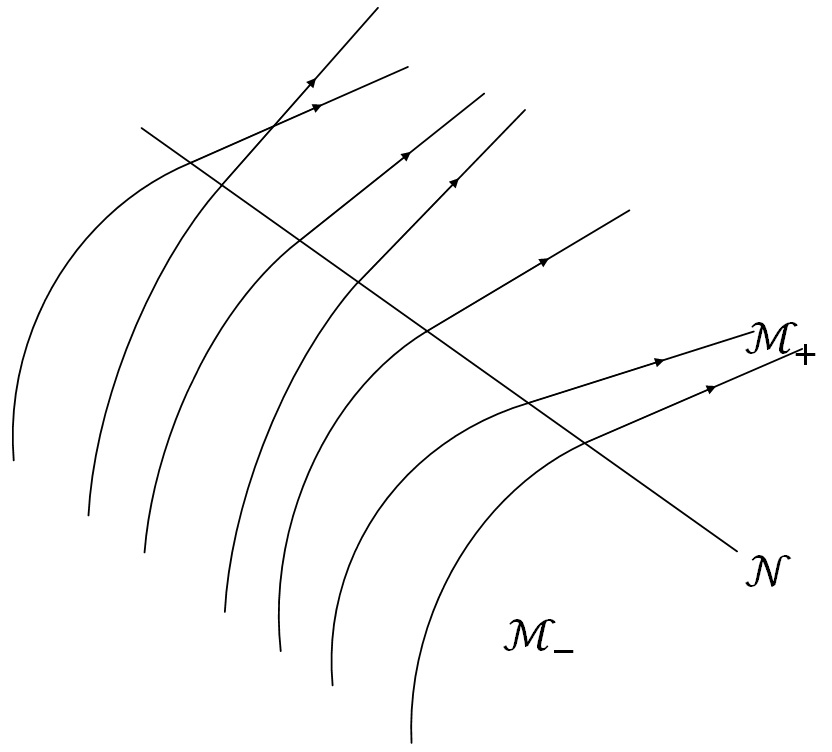}}
	\caption{In continuous coordinates the geodesic vector field is continuous across $\pazocal{N}$. Here we see that the transformed vector field to the past of $\pazocal{N}$ provides the initial conditions for the field to the future and thus the full solution to the geodesic equation.}
	\label{cts-lines}
\end{figure}

To study the evolution of a null geodesic congruence upon crossing a null shell we start directly from the geodesic equation. In continuous coordinates by definition the metric is continuous across $\pazocal N$ while the Christoffel symbols are discontinuous, and the Riemann tensor has a delta function singularity localised on the shell, these properties being directly related to the stress-energy tensor of the shell and explained in detail in \cite{Blau:2015nee}. For the purposes of our calculations we will obtain continuous coordinates across the shell by performing a coordinate transformation on $\pazocal M^-$ while leaving $\pazocal M^+$ in flat coordinates. 

\begin{figure}[h!]
	\center{\includegraphics[width=7cm]{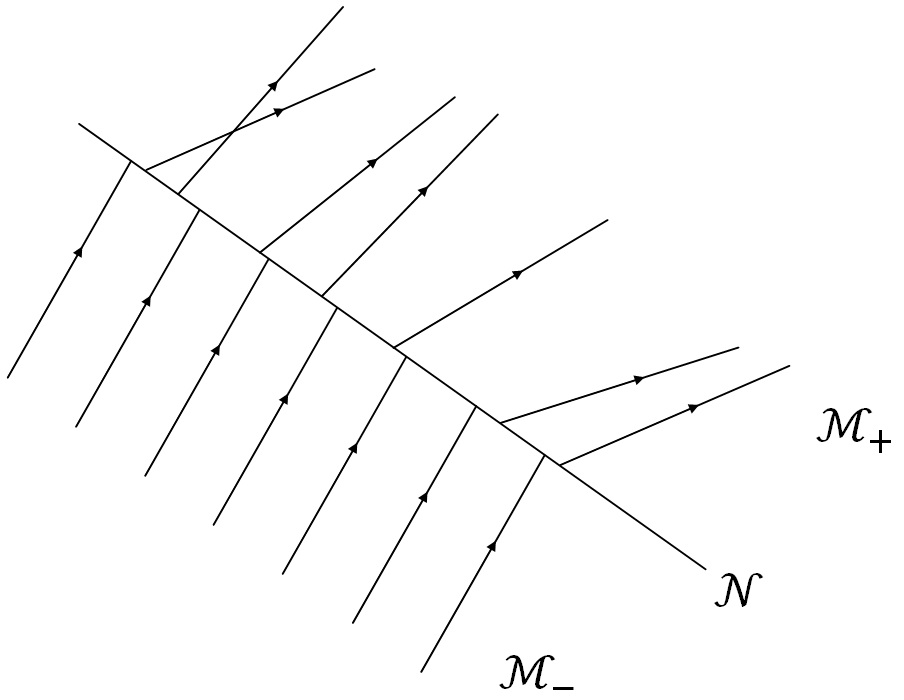}}
	\caption{With flat coordinates to the past and future the soldering transformation leads to a discontinuity across ${\pazocal N}$ in both coordinates and in the geodesic congruence.}
	\label{discts-lines}
\end{figure}

The geodesic equation in the vicinity of the shell is
\be
\ddot{X}^\mu + (\Theta(-\Phi)\Gamma^{-\mu}_{\nu\lambda}+\Theta(\Phi)\Gamma^{+\mu}_{\nu\lambda})\dot{X}^\nu\dot{X}^\lambda =0
\ee
where $\Theta$ is the Heaviside step function. It is clear that non-trivial solutions to this equation may have a discontinuity in the acceleration, but not in the tangent vector $T=\dot{X}$, and thus the geodesic flow lines are $C^1$ across the shell as shown in figure \ref{cts-lines}. Mathematically speaking this means that the geodesic vector on the shell ($T_0$) is uniquely defined $T_0=T_{\pm}\big|_{\pazocal N}$. Taking this into account we state that if the test particle has approached the hypersurface from the past then the action of crossing the hypersurface is mathematically equivalent to making a coordinate transformation on the geodesic vector from the past flat coordinates, where for the purposes of our calculations we consider a trivial constant and parallel null congruence, to the continuous coordinate system. This transformed congruence then forms the initial conditions for the congruence to the future of the hypersurface.  

\be
\label{eq:contin}
T_{+}^\alpha\big|_{\pazocal N}=\left.\left(\frac{\partial x_+^\alpha}{\partial x_-^\beta}T_-^\beta\right)\right|_{\pazocal N}
\ee

Here $T_{-}^\alpha$ is the geodesic vector of the test particle in the past domain in past flat coordinates and $T_{+}^\alpha$ is the corresponding vector after the particle crosses the shell in future coordinates as shown in both figures \ref{cts-lines} and \ref{discts-lines}. Here we should recall that all the information regarding the stress-energy tensor on the shell, which also means the effect that the shell will have on the congruence, is fully encoded in the definition of the soldering conditions and thus in the Jacobian of the soldering transformation.

Note that the geodesics are straight lines in the future and the past in the corresponding coordinate systems and with affine parameters $\lambda_{\pm}$ they are given by
\be
	x^\alpha_{\pm} = x_0^\alpha +  \lambda_{\pm} T_{\pm}^\alpha\big|_{\pazocal N}.
\ee
There is a one parameter freedom in the choice of affine parameters
\be
	\lambda_{\pm}\rightarrow\alpha^{-1}_{\pm} \lambda_{\pm},
	\qquad T_{\pm}^\alpha\big|_{\pazocal N} \rightarrow \alpha_{\pm} T_{\pm}^\alpha\big|_{\pazocal N} ,
\ee 
and the continuity equation \eqref{eq:contin} establishes a one-to-one relation between $\alpha_{-}$ and $\alpha_{+}$, thus fixing the affine parameter in the future we also fix the affine parameter in the past.


\subsection{Null Congruences crossing horizon shells}
\label{subsec:application}
Applying the proposed algorithm of the previous section we consider the
congruence $T_- = \alpha\partial_{u_-}$ globally to the past of ${\pazocal N}$ ($\alpha$ will be fixed after fixing the affine parameter to the future, as discussed in the previous section) and perform on ${\pazocal M^+}$ a coordinate transformation parametrised by $F(x^a)$ where $a=v,x,y$,
\begin{align}
\begin{split}
u_- &= \frac{u}{F_v}, \qquad \\
v_-&= F + \frac{u}{2F_v}(F_x^2 + F_y^2),\qquad \\
x_- &= x + \frac{uF_x}{F_v},\qquad \\
y_- &= y + \frac{uF_y}{F_v}.
\end{split}
\end{align}
We will refer to this transformation as a Newman-Unti soldering being the extension to a soldering of the Newman-Unti transformation $v_- = F$.        
\begin{figure}[h!]
	\center{\includegraphics[width=7cm]{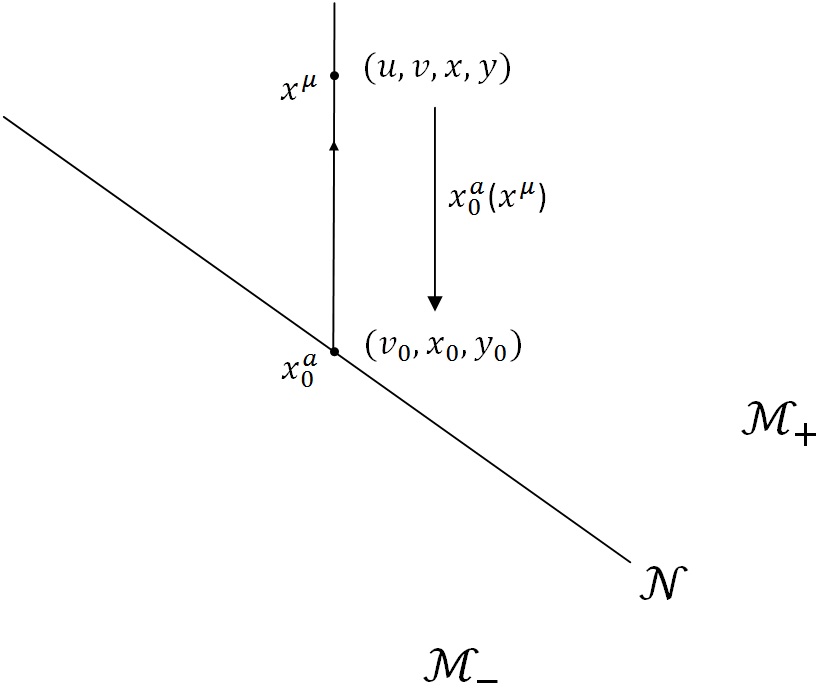}}
	\caption{Every point to the future of ${\pazocal N}$, apart from caustic points, has a unique mapping onto ${\pazocal N}$ obtained by following the geodesic of the congruence in ${\pazocal M^+}$ that passes through that point back to ${\pazocal N}$.}
	\label{maptoN}
\end{figure}
This one sided soldering transformation creates a shell at the location of  ${\pazocal N}$ and the properties of the shell are encoded in the function $F(x^a)$ as described in detail in \cite{Blau:2015nee}. To the future of ${\pazocal N}$ we have coordinates $x_+^\alpha = x^\alpha$ and we identify the future and past coordinates on ${\pazocal N}$. After the transformation the metric to the past of the shell is
\be
\label{eq:past_metric}
ds^2_-=-2\diff u \diff v +dx^2+dy^2+u\left(\frac{2}{F_v}F_{ab}dx^a dx^b\right)+\frac{u^2}{F_v^2}\left(F_{xa}F_{xb}+F_{ya}F_{yb}\right)dx^a dx^b,
\ee
while to the future it remains
\be
\label{eq:future_metric}
\diff s_+^2=-2\diff u \diff v+\diff x^2+\diff y^2.
\ee

We are interested in the value of the congruence $T_0$ on  ${\pazocal N}$ in continuous coordinates,
\be
T_0^\alpha=\frac{\partial x^\alpha}{\partial x_-^\beta}T_-^\beta\big|_{\pazocal N}
\ee
Inverting the Jacobian matrix of the coordinate transformation evaluated on ${\pazocal N}$ we find, for our choice of $T_-$, that 
\be
T_0(x_0^a) = \alpha\left.\left(F_v\partial_u + \frac{1}{2F_v}(F_x^2 + F_y^2)\partial_v - F_x\partial_x 
-F_y\partial_y\right)\right|_{\pazocal N}
\ee

The null congruences to the future of ${\pazocal N}$ are labeled by the point $(x_0^a)$ at which they cross ${\pazocal N}$ and the affine parameter $u$. Taking $T_0$ as the initial condition for the congruence on  ${\pazocal N}$ at $u=u_0=0$ we find that the null congruence to the future is described by the lines
\be
\label{eq:lines}
x^\alpha = x_0^\alpha +  u T^\alpha_0(x_0^a)
\ee
and from the $u$ component of this equation we find that
\be
	\alpha = 1/F_v.
\ee

The remaining components of \eqref{eq:lines} can in principal be inverted (in practice there will be unavoidable problems of caustics meaning that the inversion from some future points will not be well defined - we will ignore these subtleties), to obtain a projection along geodesic lines from ${\pazocal M^+}$ to ${\pazocal N}$ of the form $x_0^a = x_0^a(x^\alpha)$ as illustrated in figure \ref{maptoN}. The congruence to the future is then simply $T^\alpha(x^\mu) = T_0^\alpha(x_0^a(x^\mu))$. In the following, by a slight abuse of notation, we will use $F$ to denote the extension of the soldering transformation $F$ to the future such that $F(x^\alpha) = F(x_0^a(x^\alpha))$. A simple and useful consequence of this construction, that one can show with the help of the $a$ components of \eqref{eq:lines}, is
\be
\frac{\partial F}{\partial x^a} = \frac{\partial F}{\partial x_0^a},
\ee
With a little further work one can show that the congruence to the future of the shell is given by
\be
\label{eq:orth_hyp}
T^\mu = -\frac{1}{F_v}\ \eta^{\mu\nu}\partial_\nu F(x_0^a(x^\alpha)),
\ee
and is thus hypersurface orthogonal as anticipated at the end of \sect{\ref{subsec:memory}}.

\subsection{How the shell modifies the congruence.}
\label{subsec:behaviour}
We now turn to the projection of the B-tensor and its behaviour upon crossing the shell as described in \sect{\ref{subsec:memory}}. A natural choice for completing the tetrad along the congruence is
\be
\label{eq:proj_vectors}
e_A = -\frac{F_A}{F_v}\partial_v + \partial_A
\ee
together with $n=\partial_v$ and the tangent vector $T$. We will also need the completeness relation
\be
e^\alpha_A e^\beta_B\delta^{AB} = \eta^{\alpha\beta} + n^\alpha T^\beta + n^\beta T^\alpha.
\ee
As already discussed a null shell produces a delta function singularity in the Riemann tensor and the physical content of the shell is encoded in the jump in the orthogonal derivatives of the metric tensor
\be
\gamma_{ab} = T^\alpha[\partial_\alpha g_{ab}] = -2\frac{F_{ab}}{F_v}|_\pazocal{N}\qquad \gamma_{u\alpha}=0.
\ee
The shell in general contains matter with stress-energy tensor
\be
S_{\alpha\beta} = \mu n_\alpha n_\beta + p g_{\alpha\beta} + 2j_{(\alpha}n_{\beta)}
\ee
with $j_\alpha = (0,j_a)$. The four independent components of the stress energy tensor are the energy density $\mu$, and the surface current $j_a$, the $v$ component of which is minus the pressure $p$
\be
\mu = -\frac{1}{16\pi}\gamma_{\alpha\beta}\eta^{\alpha\beta} \qquad j_a =\frac1{16\pi}\gamma_{a\beta} n^\beta\qquad p = -j_v =  -\frac{1}{16\pi}\gamma_{\alpha\beta}n^\alpha n^\beta.
\ee
These account for four out of the six independent components of $\gamma_{\alpha\beta}$, the remaining two coming from the spatial $(x,y)$ part of $\hat{\gamma}_{\alpha\beta}$
\be
\hat{\gamma}_{\alpha\beta} = \gamma_{\alpha\beta} - \frac12\gamma_{\delta\kappa}\eta^{\delta\kappa}\eta_{\alpha\beta}
\ee
which contribute to the Weyl tensor and encode the two polarisations of an impulsive gravitational wave on the shell. We will see in detail how this works below. 

To study the behaviour of a null congruence crossing the null shell we need to calculate $R_{AB}$ and it is straightforward to show that
\be
\bar{R}_{AB} = -\frac12\gamma_{\alpha\beta}e_A^\alpha e_B^\beta = -\frac12\gamma_{AB}.
\ee

Given the Einstein equation
\be
R_{\mu\nu} - \frac12 g_{\mu\nu}R = 8\pi S_{\mu\nu}\delta(u)
\ee
we can relate the trace of $\bar{R}_{AB}$ to the surface quantities
\be
\bar{\mathcal{R}} = 8\pi S_{\mu\nu}T^\mu T^\nu = 8\pi\mu - 16\pi j_aT^a,
\ee
while the projection of the Weyl tensor on the congruence is
\be
\bar{C}_{AB} = -\frac12\gamma_{AB} + \frac14\gamma^C_C\delta_{AB} = -\frac{1}{2}\hat{\gamma}_{\alpha\beta}e_A^\alpha e_B^\beta + 16\pi j_a T^a\delta_{AB}.
\ee

\subsubsection{Newman-Unti soldering transformations}

Taking the explicit form for $T^\mu$ from the previous section we find for a general Newman-Unti type transformation that
\be
\label{eq:general_B}
\pazocal{B}_{AB} = e_A^\alpha e_B^\beta \pazocal{B}_{\alpha\beta} = -\frac{F_{AB}}{F_v} - F_AF_B\frac{F_{vv}}{F_v^3} + \frac{(F_A F_{Bv} + F_B F_{Av})}{F_v^2}.
\ee
Evaluating $\pazocal{B}_{AB}$ on the shell gives us directly its discontinuity given that we have taken a congruence with $\pazocal{B}_{AB}=0$ before the shell. In this expression we must take care to recall that although $\partial_{a0}F = \partial_aF$ second derivatives must include the Jacobian of the mapping $x_0^a(x^\alpha)$. We see that $\pazocal{B}_{AB}$ is symmetric and thus the congruence has zero rotation consistent with the hypersurface orthogonality demonstrated in the previous section and also the more general arguments of \sect{\ref{subsec:memory}}.

Evaluating explicitly $\bar{\mathcal R}$ and $\bar{C}_{AB}$ and comparing to \eqref{eq:general_B} we find that the change in expansion upon crossing the shell
\be
\label{eq:dtheta}
\theta|_\pazocal{N} = -\bar{\mathcal R} = - 8\pi\mu + 16\pi j_aT^a
\ee
is determined by a combination of the shell energy density and surface currents while the change in the shear
\be
\label{eq:dC}
{\sigma_{AB}}|_\pazocal{N} = -\bar{C}_{AB} = \frac{1}{2}\hat{\gamma}_{\alpha\beta}e_A^\alpha e_B^\beta - 16\pi j_a T^a\delta_{AB}
\ee
is determined by the gravitational wave component and surface current of the shell.

\subsubsection{BMS soldering}

To explicitly evaluate $\pazocal{B}_{AB}$ \eqref{eq:general_B} also to the future of the shell we need to invert equations \eqref{eq:lines} as discussed in the previous section. We will simplify the following calculations by just considering the special case of BMS supertranslation solderings and thus we take 
\be
\label{eq:bms_soldering}
F(v,x,y) = v + f(x,y).
\ee

Then
\be
\label{eq:bms_geodesic}
T_{\alpha} = -\partial_\alpha F = (-\frac12(f_x^2 + f_y^2),-1, -f_x,-f_y)
\ee
and
\be
\pazocal{B}_{AB} =  -\partial_B f_A = - \frac{\partial x_0^C}{\partial x^B}\frac{\partial f_A}{\partial x_0^C}.
\ee
In this case we need only the Jacobian of the transformation on spatial coordinates that we obtain by taking derivatives of the $x,y$ components of \eqref{eq:lines} with respect to $x_A = (x,y)$ to obtain
\be
\delta_A^B = \frac{\partial x_0^C}{\partial x^A}(\delta_C^B - u f_{BC})
\ee
and inverting we find the Jacobian of the transformation
\be
\left(\frac{\partial x_0^B}{\partial x^A}\right) = \frac{1}{1 - u\trf + u^2\detf}
\begin{pmatrix}
	1-uf_{yy}&uf_{xy} \\
	uf_{xy} &1-uf_{xx} 
\end{pmatrix},
\ee
where  $\trf=f_{xx}+f_{yy}$ and $\detf = f_{xx}f_{yy} - f_{xy}^2$.
Thus
\be
	\pazocal{B} =\frac{-1}{1 - u\trf + u^2\detf}
	\begin{pmatrix}
		f_{xx} - u\detf &f_{xy} \\
		f_{xy} & f_{yy} - u\detf
	\end{pmatrix}
\ee
corresponding to the expansion 
\be
\label{eq:expansion} 
\theta =\frac{-\trf + 2 u\,\detf}{1 - u \trf + u^2 \detf}
\ee
and shear
\be
\label{eq:shear}
\sigma =\frac{-1}{2(1 - u \trf + u^2 \detf)} 
\begin{pmatrix}
	f_{xx}-f_{yy}&f_{xy} \\
	f_{xy}& -f_{xx}+f_{yy}
\end{pmatrix}.
\ee
Evaluating
\be
\bar{\mathcal R} = f_{xx} + f_{yy} = 8\pi\mu\qquad
\bar{C} = \frac12\begin{pmatrix}
	f_{xx}- f_{yy}&2f_{xy} \\
	2f_{xy} & -f_{xx}+f_{yy}
\end{pmatrix} = -\frac12\hat{\gamma}
\ee
it is easy to check that our solutions for expansion and shear on and to the future of $\pazocal{N}$ satisfy the evolution equations
\be
\frac{\dd \theta}{\dd u} = - \frac12 \theta^2 - 2(\sigma_+^2 + \sigma_\times^2) - 8\pi\mu\delta(u)
\qquad
\text{and}
\qquad
\frac{\dd \sigma}{\dd u} = -\theta  \sigma +\frac12\hat{\gamma}\delta(u).
\ee

We see in particular that for the BMS transformations the $\pazocal{B}$-memory effect corresponds to a jump in the expansion upon crossing the shell that is proportional to the energy density of the shell together with a change in the shear that is proportional to the gravitational wave component of the shell. 

\subsection{Discussion}
\label{subsec:discussion}

We have presented a new approach for studying congruences that cross a singular hypersurface. Our method is based on the physically justified assumption that the geodesic vector of a test particle is continuous across the hypersurface when using continuous coordinates. To obtain the geodesic flow to the future of the hypersurface one simply needs to do a coordinate transformation on the past coordinates to go to a continuous coordinate system. The resulting transformation on the geodesic congruence in $\pazocal M^-$ gives initial conditions on $\pazocal N$ to develop the geodesic vector field on $\pazocal M^+$ to the future.

We then proved that a parallel congruence upon crossing the shell gives rise to a hypersurface orthogonal congruence to the future of the shell, and in particular that the shell gives rise to a discontinuity in the $\pazocal{B}$-tensor of the congruence. In general the jump in the expansion is determined by the energy density and currents on the shell while the jump in the shear is determined by the gravitational wave component together with the surface currents. Although we derived these results using a particular congruence, it should be clear from \eqref{eq:dtheta} and \eqref{eq:dC} that the results are independent of the choice of congruence in the case of BMS supertranslations for which the surface currents are zero. We also provide a general argument that a hypersurface orthogonal congruence before the shell will give rise to a hypersurface orthogonal congruence to the future.

The change in the $\pazocal{B}$-tensor after the passage of an outgoing gravitational wave leads to a covariant description of the gravitational memory effect - the $\pazocal{B}$-memory effect.
Although our construction and approach to gravitational memory appears to be quite distinct from that reviewed in \cite{Ashtekar:2018lor} there are many intriguing similarities. They introduce a trace free ``shear like'' tensor
$\sigma_{ab} = \nabla_a\nabla_b f$ where $f$ is the shift in a BMS supertranslation
on $\pazocal{I}$ and the Lie derivative along $\pazocal{I}$ of $\sigma_{ab}$ is the news tensor $N_{ab}$. The picture that emerges suggests that the outgoing null shell induces a BMS supertranslation on $\pazocal{I}$ in the same way that a soft graviton is supposed to \cite{Strominger:2017zoo}. 

It would be very interesting to study the quantum version of this effect and the calculation of the eikonal wavefunction may be a first step in such an approach. In the eikonal picture the local wavefronts of a wavefunction follow the geodesics of the spacetime. The presence of an outgoing gravitational wave produces a radical reorganization of the congruence such that in general a flat wavefront can be distorted in a myriad of different ways. One may imagine that at a deeper level this distortion corresponds to a radical change in the quantum field theory vacuum that is constructed from plane wave states. It would be interesting in particular to investigate how the propagation across the shell of a good basis of wave-functions may not give rise to a reasonable basis to the future of the shell given that BMS transformations map between inequivalent quantum field theory vacuum states \cite{Ashtekar:2018lor}.

%% file: Integrability_of_MP_1.tex
\section[Integrability of geodesics in near-horizon extremal Myers-Perry black holes: Special cases]
		{Integrability of geodesics \\ in near-horizon extremal Myers-Perry \\ black holes: Special cases}
\label{sec:Integrability_of_MP_1}
Any dynamical system, particle or field dynamics alike, is classically described by equations of motion and some boundary conditions for the field theory case. The main task in analyzing the system is to solve the equations of motion, which are generically (partial) second order differential equations, and solving them  is generically a formidable task. Symmetries, Noether theorem and constants of motion, are the usual tools facilitating tackling the problem.   In this and the following chapter we will focus on particle dynamics on certain $d$ dimensional curved backgrounds.
 
In a dynamical system with $N$ degrees of freedom and hence a $2N$ dimensional phase space, if  number of  independent symmetries is equal to $N$,  the system is called \emph{integrable} and is usually solvable. If the system possesses  $N+p,$ $1\leq p\leq N-1$, independent symmetries (and hence functionally independent constants of motion),  it is called \emph{superintegrable} and the region it can probe in its $2N$ dimensional phase space is a compact $N-p$  dimensional surface; e.g. see \cite{Kupershmidt:1990ws, Miller:2013gxa, oxana}.

For the question of particle dynamics on a general curved (usually a black hole) background in $d$ dimensions, we are dealing with a $2d$ dimensional phase space. It is an established fact that isometries of the background, the Killing vectors, provide a set of constants of motion. Moreover, reparametrization invariance of the particle action implies that there is always a second rank Killing-tensor whose conserved charge is the mass of particle. For backgrounds of interest, e.g. black holes or their near horizon geometries, usually the number of Killing vectors plus one is less than $d$ and one may wonder if the system is integrable.

The question of integrability of particle dynamics on black hole or near horizon geometries have been extensively analyzed in the literature e.g. see \cite{Carter, conformal-mechanics-BH-1,conformal-mechanics-BH-2_1,conformal-mechanics-BH-2_2,conformal-mechanics-BH-2_3,conformal-mechanics-BH-2_4,conformal-mechanics-BH-2_5,conformal-mechanics-BH-2_6,conformal-mechanics-BH-2_7,conformal-mechanics-BH-2_8,GNS-11,GNS-12, Frolov:2003en, Benenti and Francaviglia, Kubiznak:2007kh,Krtous:2006qy,Frolov:2006pe,Cariglia:2011qb,Hidden-symmetry-NHEK, Lunin}. In particular, it has been shown that the problem is (super)integrable for a large class of black holes. The integrability is often associated with the existence of higher rank, usually second rank, Killing tensor fields \cite{Carter} (see \cite{Frolov:2017kze} for review).

Given an extremal black hole there are general theorems stating that in the near horizon limit we obtain a usually smooth geometry with larger isometry group than the original extremal black hole \cite{NHEG-general}. It is hence an interesting question to explore if this symmetry enhancement yields further independent constants of motion and how it affects the (super)integrability of particle dynamics.
This question, besides the academic interests, is also relevant to some of the observations related to black holes: It is now a well-accepted fact that there are fast rotating black holes in the sky which are well modeled by an extreme Kerr geometry \cite{Extreme-Kerr} and the matter moving around these black holes in their accretion disks are essentially probing the near horizon geometry \cite{Falcke:1999pj}.

{The isometry group of generic stationary extremal black holes in the near horizon  region is shown to have an $SO(2,1)=SL(2,\mathbb{R})$ part \cite{NHEG-general,NHEG-2}. Therefore, particle dynamics on the near horizon extreme geometries possesses  dynamical $0+1$ dimensional conformal symmetry, i.e.  it defines a ``conformal mechanics" \cite{conformal-mechanics-BH-1,conformal-mechanics-BH-2_1,conformal-mechanics-BH-2_2,conformal-mechanics-BH-2_3,conformal-mechanics-BH-2_4,conformal-mechanics-BH-2_5,conformal-mechanics-BH-2_6,conformal-mechanics-BH-2_7,conformal-mechanics-BH-2_8,GNS-11,GNS-12}. This allows to reduce the problem to the study of  system depending on
latitudinal and azimuthal coordinates and their conjugate momenta with the effective Hamiltonian being Casimir of conformal algebra. Such associated systems have been investigated from  various viewpoints  in Refs. \cite{Armen-Tigran1,Armen-Tigran2,Armen-Tigran3,Armen-Tigran4,Armen-Tigran5} where they were called ``angular (or spherical) mechanics''.
	
In this work, we continue our analysis of \cite{non-equal-general, Demirchian:2017uvo} and extend the analysis there to Near Horizon Extremal Myers-Perry \cite{mp} (NHEMP) black holes \cite{NHEG-2} in general odd and even dimensions. We discuss the separability of variables, constants of motion  for ``angular mechanics" associated with
these systems and how they are related to the second rank Killing tensors of the background. } While the system is in general integrable, as we show, there are special cases where the system is superintegrable. Moreover, we discuss another interesting case, the Extremal Vanishing Horizon (EVH)  \cite{NHEVH-1} Myer-Perry black holes \cite{NHEVH-MP} and  show the integrability of geodesics in the Near Horizon EVH Myers-Perry (NHEVH-MP) geometries.

In \sect{\ref{NHEMP-background-Sec}} we present the geometry of near-horizon extremal Myers-Perry black holes in generic even and odd dimensions, and  construct the ``angular mechanics" describing  probe particle dynamics. In this section we set our notations and conventions. 

In \sect{\ref{Particle-dynamics-Sec}} we analyze generic causal curve, massive or massless geodesic, in the NHEMP background. We show that this Hamiltonian system is separable in ellipsoidal coordinate system, work out the constants of motion and establish that the system is integrable. Moreover, we show how the Killing vectors and second rank Killing tensors are related to these constants of motion. In \sect{\ref{Special-cases-Sec}} we analyze special cases where some of the rotation parameters of the background NHEMP are equal. In these cases we have some extra Killing vectors and tensors and the system is superintegrable. \sect{\ref{EVH-Sec}} contains the analysis of particle dynamics on the special class of Extremal Vanishing Horizon (EVH) Myers-Perry black holes. We end this note with discussions and further  comments.

This chapter was based on the papers \cite{Demirchian:2017uvo,Demirchian:2018xsk,Demirchian_PAN,Demirchian_PPN}

\subsection{NHEMP in arbitrary dimensions; unified description}\label{NHEMP-background-Sec}

The NHEMP metric in both odd and even dimensions in the  Gaussian null coordinates was presented in \cite{NHEG-2}. The NHEMP is a (generically) a smooth solution to vacuum Einstein equations in odd $d=(2N+1)$- and $d=(2N+2)$-dimensions, in general it is specified by $N$ number of rotation parameters $a_i$ (or $N$ angular momenta $J_i$) and has $SL(2,\mathbb{R})\times U(1)^N$ isometry. In Boyer-Lindquist coordinates NHEMP metric has the form
\begin{gather}\label{NHEMP-metric-rt-coord}
ds^2=\frac{F_H}{b}\left(-r^2d\tau^2+\frac{dr^2}{r^2}\right)+\sum_{I=1}^{N_\sigma}(r_H^2+a_I^2)d\mu_I^2+\gamma_{ij}D\varphi^i D\varphi^j,
\\
D\varphi^i\equiv d\varphi^i+\frac{B^i}{b}rd\tau,
\end{gather}
where 
$N_\sigma=[\frac{d}{2}]=N+\sigma$, i.e. $\sigma=0$ for the odd and $\sigma=1$ for the even dimensions cases,
$r_H$ is a black hole radius
which satisfy the equation
\be
	\sum_{I=1}^{N_\sigma}\frac{r^2_H}{r^2_H+a^2_I}=\frac{1+2\sigma}{1+\sigma}, \qquad{\rm with}\quad a_{N+1}=0
\label{ai}\ee
and,\footnote{There seems to be a minor typo in the exressions for NHEMP metrics given in \cite{NHEG-2}, which we have corrected here.}
\begin{gather}
F_H=1-\sum_{i=1}^{N}\frac{a_i^2\mu_i^2}{r_H^2+a_i^2},
\qquad	B^i=\frac{2r_Ha_i}{(r_H^2+a_i^2)^2},
\end{gather}
\begin{gather}
 b=\frac{1}{r_H^2}\left(\sum_{i=1}^{N}\frac{\sigma\, r_H^2}{r_H^2+a_i^2}+4\sum_{i<j}^{N}\frac{r_H^2}{r_H^2+a_i^2}\frac{r_H^2}{r_H^2+a_j^2}\right),
\end{gather}
\begin{gather}
\gamma_{ij}=(r_H^2+a_i^2)\mu_i^2\delta_{ij}+\frac{1}{F_H}a_i\mu_i^2a_j\mu_j^2,
\qquad \sum_{I=1}^{N_\sigma}\mu_I^2=1.
\end{gather}
In our notations  lowercase Latin indices $i, j$  run from $1$ to $N$ and uppercase Latin indices $I,J$ run over
1 to  $N_\sigma$.

For the case when all $a_i$ take generic non-zero values\footnote{The case when one of the $a_i$ is zero is the EVH case we will discuss separately in \sect{\ref{EVH-Sec}}.} it is convenient to introduce
new parameters $m_i$
\be
m_i=\frac{r_H^2+a_i^2}{r_H^2} > 1,\qquad m_{N+1}=1\;\quad  \text{and} \quad \sum^{N_\sigma}_{I=1}\frac{1}{m_I}= \frac{1+2\sigma}{1+\sigma},
\ee
and re-scaled coordinates $x_I$,
\be
x_I=\sqrt{m_I}\mu_I \;:\quad \sum_{I=1}^{N_\sigma}\frac{x^2_I}{m_I}=1.
\ee
In these terms the near-horizon metrics reads
\be
\begin{gathered}
\label{m2}
	\frac{ds^2}{r^2_H}=A(x)\left(-r^2d\tau^2+\frac{dr^2}{r^2}\right)+\sum_{I=1}^{N_\sigma} dx_Idx_I+
	\sum_{i,j=1}^N\tilde{\gamma}_{ij}x_i x_j D\varphi^iD\varphi^j, \\   
	D\varphi^i\equiv d\varphi^i+k^ird\tau,
\end{gathered}
\ee
where
\be\label{gamma}
\begin{gathered}
	A(x) ={\frac{\sum_{I=1}^{N_\sigma} x^2_I/m^2_I}{\frac{\sigma}{1+\sigma}+4\sum_{i<j}^{N}\frac{1}{m_i}\frac{1}{m_j}}}, \quad
	\tilde{\gamma}_{ij}=\delta_{ij}+ \frac{1}{\sum_I^{N_\sigma} x_I^2/m^2_I}\frac{\sqrt{m_i-1}x_i}{m_i}  \frac{\sqrt{m_j-1}x_j}{m_j},\\
	k^i=\frac{2\sqrt{m_i-1}}{m^2_i(\frac{\sigma}{1+\sigma}+4\sum_{k<l}^{N}\frac{1}{m_k}\frac{1}{m_l})},
\end{gathered}
\ee
with
\be\label{eq:restr_on_x}
\sum_{I=1}^{N_\sigma} \frac{x^2_I}{m_I}=1, \qquad \sum_{I=1}^{N_\sigma} \frac{1}{m_I}=\frac{1+2\sigma}{1+\sigma}.
\ee
With this unified description at hands we are ready to describe probe particle dynamics.

\subsection{Probe-particle dynamics}

The metric \eqref{m2} has $SL(2,\mathbb{R})$ isometry group and hence  the particle dynamics on this background  exhibits  dynamical conformal symmetry; we are dealing with a ``conformal mechanics" problem \cite{conformal-mechanics-BH-1,conformal-mechanics-BH-2_1,conformal-mechanics-BH-2_2,conformal-mechanics-BH-2_3,conformal-mechanics-BH-2_4,conformal-mechanics-BH-2_5,conformal-mechanics-BH-2_6,conformal-mechanics-BH-2_7,conformal-mechanics-BH-2_8, GNS-11,GNS-12}. Let us denote the three generators of this $sl(2,\mathbb{R})$ algebra by $H,D,K$, and its Casimir by ${\cal I}$:
\be
\{H,D\}=H, \quad \{H,K \}=2D, \quad \{D,K \} =K,\qquad \mathcal{I}=HK-D^2.
\label{confalg}\ee

The mass-shell equation for a particle of mass $m_0$ moving in the background metric
\be\label{mass-shell-1}
m^2_0=-\sum_{A,B=1}^{2N+1+\sigma}g^{AB}p_A p_B,
\ee
leads to the following expression
\be
\label{eq:mass_shell} 			m_0^2r_H^2=\frac{1}{A}\left[\left(\frac{p_0}{r}-\sum_{i=1}^{N}k_ip_{\varphi_i}\right)^2-(rp_r)^2\right]-\sum_{a,b=1}^{N_\sigma-1}h^{ab}p_ap_b
-\sum_{i,j=1}^{N}\tilde{\gamma}^{ij}\frac{p_{\varphi_i}}{x_i}\frac{p_{\varphi_j}}{x_j},
\ee
where
\be
\begin{gathered}
	h^{ab}=\delta^{ab}-\frac{1}{\sum\limits_{I=1}^{N_\sigma} x^2_I/m^2_I}\frac{x_a}{m_a}\frac{x_b}{m_b}, \qquad a,b=1,\cdots N_\sigma-1,
	\\
	\tilde{\gamma}^{ij}=\delta^{ij}-x_i\frac{\sqrt{m_i-1}}{m_i}x_j\frac{\sqrt{m_j-1}}{m_j},\qquad i,j=1,\cdots, N.
\end{gathered}
\ee
Using \eqref{eq:mass_shell}, as in \cite{non-equal-general}, we can construct the Hamiltonian $H=p_0$ and the other generators of the conformal algebra
\begin{gather}
\label{eq:Hamiltonian_initial}
H= r\left(\sqrt{ L(x_a, p_a, p_{\varphi_i}) +(rp_r)^2}+\sum_{i=1}^N k_ip_{\varphi_i}\right),\\
D=r p_r,\qquad
K=\frac{1}{r} \left(\sqrt{ L(x_a, p_a, p_{\varphi_i}) +(rp_r)^2}- \sum_{i=1}^N k_ip_{\varphi_i}
\right),
\end{gather}
where
$$
L(x_a, p_a, p_{\varphi_i})= A\left(m_0 r_H^2+ \sum_{a,b=1}^{N_\sigma-1}h^{ab}p_ap_b
+\sum_{i,j=1}^{N}\tilde{\gamma}^{ij}\frac{p_{\varphi_i}}{x_i}\frac{p_{\varphi_j}}{x_j} \right),
$$
and the momenta $p_a, p_{\varphi_i}, p_r$ are conjugate to $x_a, \varphi_i, r$ with the canonical Poisson brackets
\be
\label{Poisson-bracket}
\{p_a,x_b\}=\delta_{ab}, \qquad \{p_{\varphi_i},\varphi_j\}=\delta_{ij},\qquad\{p_r, r\}=1.
\ee
Thus, the   Casimir element of the conformal algebra  reads
\be
\label{L}
\mathcal{I}=A\left[\sum_{a,b=1}^{N_\sigma-1}h^{ab}p_ap_b+\sum_{i=1}^{N}\frac{p_{\varphi_i}^2}{x_i^2}+g_0\right]- \mathcal{I}_0
\ee
where
\be
\label{consts}
g_0=-\left(\sum_{i=1}^N \frac{\sqrt{m_i-1}p_{\varphi_i}}{m_i}\right)^2+m^2_0r^2_H,
\qquad
\mathcal{I}_0=\left(\sum_i^N k_ip_{\varphi_i}\right)^2.
\ee
In an appropriately chosen frame $H$ can be written in formally nonrelativistic form
\cite{conformal-mechanics-BH-2_1,conformal-mechanics-BH-2_2,conformal-mechanics-BH-2_3,conformal-mechanics-BH-2_4,conformal-mechanics-BH-2_5,conformal-mechanics-BH-2_6,conformal-mechanics-BH-2_7,conformal-mechanics-BH-2_8,GNS-11,GNS-12}
\be\label{nonrel}
H=\frac{p^2_R}{2}+\frac{2\mathcal{I}}{R^2},
\ee
where $R=\sqrt{2K}$, $ p_R=\frac{2D}{\sqrt{2K}}$ are  the effective ``radius" and its canonical conjugate ``radial momentum''.
As we will show below the Casimir  $\mathcal{I}$  encodes all the essential information about the system of particle on these backgrounds.
The Casimir   $\mathcal{I}$ \eqref{L} is at most quadratic in momenta canonically conjugate to the remaining angular variables and it can conveniently be viewed as the Hamiltonian of a reduced ``angular/spherical mechanics" \cite{Armen-Tigran1,Armen-Tigran2,Armen-Tigran3,Armen-Tigran4,Armen-Tigran5}  describing motion of particle on some curved background.
Note that the  ``time parameter''    conjugate  to $\mathcal{I}$  is different   than the time  parameter $\tau$ appeared in metric \eqref{m2} whose conjugate variable is $H=p_0$. 
See  \cite{SIGMA} for more detailed discussions.

Since  the azimuthal angular variables $\varphi^i$ are cyclic, corresponding conjugate momenta $p_{\varphi_i}$ are constants of motion.
We then remain with a reduced $(N_\sigma-1)$-dimensional system described by Hamiltonian \eqref{L} and $x_a$ variables and their conjugate momenta.

\subsection{Fully Non-isotropic case}\label{Particle-dynamics-Sec}


To show that the angular/spherical mechanics system is integrable, we show that it is separable in the ellipsoidal coordinates
when  we are dealing with cases where all parameters $m_i$ are non-equal.
The ellipsoidal coordinates $\lambda_I$ for odd and even dimensions are then defined as
\be
\label{xN}
x^2_I=(m_I-\lambda_I)\prod_{J=1, J\neq I}^{N_\sigma}\frac{m_I-\lambda_J}{m_I-m_J},\qquad \lambda_{N_\sigma}  < m_{N_\sigma}  <  \ldots < \lambda_2  < m_2  < \lambda_1 < m_1.
\ee
To resolve the condition $\sum_{I=1}^{N_\sigma} \frac{x^2_I}{m_I}=1$ we choose $\lambda_{N_\sigma}=0$ and hence there are $N_\sigma-1$ independent $\lambda_I$ variables, which will be denoted by $\lambda_a$.

In these coordinates the  angular Hamiltonian $\mathcal{I}$ (shifted by a constant and appropriately rescaled) reads
\be\label{eq:conf_ham_odd}
\tilde{\mathcal{I}}=\lambda_1\ldots\lambda_{N_\sigma -1}\left[ - \sum_a^{N_\sigma-1}\frac{{4\prod_{I=1}^{N_\sigma}(m_I-\lambda_a) }\pi^2_a}{\lambda_a\prod_{b=1,a\ne b}^{N_\sigma-1}(\lambda_b-\lambda_a)}+\sum_{i=1}^{N_\sigma}\frac{{g}^2_I}{\prod_{a=1}^{N_\sigma-1}(m_I-\lambda_a)} +g_0\right],
\ee
where
\be	
\tilde{\mathcal{I}}\equiv  \left({\mathcal{I}+\mathcal{I}_0}\right)\left(\frac{\sigma}{1+\sigma}+4\sum_{k<l}^{N}\frac{1}{m_k}\frac{1}{m_l}\right)\prod_{i=1}^{N}m_i,\qquad \mathcal{I}_0=\left(\sum_i^N k_ip_{\varphi_i}\right)^2,
\ee
with
\be
{g}^2_I =\frac{p^2_{\varphi_I}}{m_I}\prod_{J=1,J\neq I}^{N_\sigma} (m_I-m_J),
\qquad	g_{N+1}=p_{\varphi_{N+1}}\equiv 0,
\ee
and $\{\pi_a,\lambda_b\}=\delta_{ab}$, $\{p_{\varphi_i},\varphi_j\}=\delta_{ij}$.

The level surface  of angular Hamiltonian \eqref {eq:conf_ham_odd}, $\tilde{\mathcal{I}}={\cal E}$,    can be conveniently represented through
\be\label{HJO}	
\sum_{a=1}^{N_\sigma -1}\frac{R_a- \mathcal{E}}{\lambda_a\prod_{b=1,a\ne b}^{N_\sigma -1}(\lambda_b-\lambda_a)}=0,
\ee
where\footnote{Note that $R_a\lambda_a\rightarrow R_a$ and $\nu_a \rightarrow F_{a+1}$ replacements have been assumed in the  current chapter compared to \cite{non-equal-general}.}
\be
R_a\equiv -4\prod_{I=1}^{N_\sigma}(m_I-\lambda_a){\pi_a^2}+(-1)^{N_\sigma}\sum_{I=1}^{N_\sigma}\frac{g_I^2\lambda_a}{m_I-\lambda_a}-g_0(-\lambda_a)^{N_\sigma-1},
\label{24}\ee
and we  used the  identities
\be
\label{26}
\begin{gathered}
	\frac{1}{\prod_{a=1}^{N_\sigma-1}(\lambda_a-\kappa)} =   \sum_{a=1}^{N_\sigma-1}\frac{1}{\prod_{b=1;a\ne b}^{N_\sigma-1}(\lambda_b-\lambda_a)}\frac{1}{\lambda_a-\kappa},\\
	\frac{1}{\lambda_1\ldots\lambda_{N_\sigma-1}}= \sum_{a=1}^{N_\sigma-1} \frac{1}{\prod_{b=1; b\neq a}^{N_\sigma-1}(\lambda_b-\lambda_a)}\frac{1}{\lambda_a}.
\end{gathered}
\ee
We can rewrite the expression \eqref{HJO} in more useful form, recalling the identities,
\be\label{eq:relation_1}
\sum_{a=1}^{N_\sigma -1}\frac{\lambda_a^{\alpha}}{\prod\limits_{\substack{b=1\\{b\ne a}}}^{N_\sigma -1}(\lambda_a-\lambda_b)}=\delta_{\alpha,N_\sigma-2} \qquad\alpha=0,...,N_\sigma-2.
\ee
Multiplying both sides of \eqref{eq:relation_1} by arbitrary constants $\nu_\alpha$ and  adding  to  \eqref{HJO}, we get
\be\label{HJ1}
\sum_{a=1}^{N_\sigma -1}\frac{R_a(\pi,\lambda)
	-\sum_{c=1}^{N_\sigma-1}\nu_{c-1}\lambda^{c-1}_a}{\lambda_a\prod_{b=1,a\ne b}^{N_\sigma -1}(\lambda_b-\lambda_a)}=0,\qquad \nu_0=\mathcal{E}.
\ee

Equipped with the above we can solve the  Hamilton-Jacobi equations
\be
{\mathcal{E}}(\lambda_a,\frac{\partial S_{gen}}{\partial \lambda_a} )=\nu_0,
\ee
and obtain the generating function $S_{gen}$ depending on $N_\sigma-1$ integration constants (i.e. the  general solution of Hamilton-Jacobi equation).
To this end we substitute in \eqref{HJ1}
\be
\pi_a=\frac{\partial S_{gen}}{\partial \lambda_a},
\ee
and  choose the ansatz
\be\label{S-gen}
S_\text{gen}(\lambda_1,\dots,\lambda_{N_\sigma-1})=\sum_{a=1}^{N_\sigma-1} S(\lambda_a).
\ee
This reduces the Hamilton-Jacobi equation to a set of $N_\sigma -1$ ordinary differential equations
\be
\label{31}
R\Big(\lambda_a,\frac{d S(\lambda_a)}{d\lambda_a}\Big)
-\sum_{b =1}^{N_\sigma -1}\nu_{b-1}\lambda^{b-1}_a=0,
\ee
or in an  explicit form,
\begin{align}
\begin{split}
-{4}\left(\frac{dS(\lambda_a)}{d\lambda_a}\right)^2\prod_{I=1}^{N_\sigma}{(m_I-\lambda_a)}
&+ (-1)^{N_\sigma}\sum_{I=1}^{N_\sigma}\frac{{ g}^2_I {\lambda_a}}{m_I-
	\lambda_a}\\
&-g_0(-\lambda_a)^{N_\sigma-1}-\sum_{b =1}^{N_\sigma -1}\nu_{b-1}\lambda^{b-1}_a=0.
\label{part}
\end{split}
\end{align}
Hence,  the analytic solution to the Hamilton-Jacobi equation is given through the generating function \eqref{S-gen}  with
\be\label{Generating-function}
S(\lambda,\nu_a )=\frac 12\frac{d\lambda}{\sqrt{
		\prod_{I=1}^{N_\sigma}{(m_I-\lambda)}}}
\sqrt{
	(-1)^{N_\sigma}\left[\sum_{I=1}^{N_\sigma}\frac{{ g}^2_Im_I}{m_I-\lambda}
	+g_0\lambda^{N_\sigma-1}-\sum_{i=1}^N{ g}^2_i\right]
	-\sum_{b =1}^{N_\sigma -1}\nu_{b-1}\lambda^{b-1}}\ .
\ee
Then, differentiating with respect to constants $\nu_a$,  we can get the explicit solutions of the equations of motion
\begin{equation}
\begin{split}
&\tau=\frac{\partial S_{gen}}{\partial \nu_0}\equiv\frac{\partial S_{gen}}{\partial {\cal E}},\\
& c_a=\frac{\partial S_{gen}}{\partial \nu_a}
\end{split}
\end{equation}

To include the dynamics of azimuthal coordinates $\varphi_i$ we have to consider the generating function
$
S_{tot}=S_{gen}+\sum_{i=1}^N p_{\varphi_i}\varphi_i,
$
where we take into account functional dependence of  $g_0, g_i$ from $p_{\varphi_i}$.
This yields the solutions for azimuthal coordinates
\be
\varphi_i=-\frac{\partial S_{gen}}{\partial p_{\varphi_i}}.
\ee
Thus, we get the  solutions of the angular sector of generic NHEMP with non-equal non-vanishing rotational parameters.\\

\subsubsection{Constants of motion }

The expressions for commuting constants of motion $F_a$ can be found from \eqref{31}, by expressing constants $\nu_a$ in terms of  $\lambda_a, \pi_a=\partial S_{gen}/\partial\lambda_a$:
\be
\begin{aligned}
\sum_{b =1}^{N_\sigma-1}F_b\lambda^{b-1}_a=R_a(\pi_a,\lambda_a)
&\qquad
\Longleftrightarrow \\
&\begin{pmatrix}
	1 & \lambda_1 & \lambda_1^2 & \cdots & \lambda_1^{N_\sigma-2}
	\\
	1 & \lambda_2 & \lambda_2^2 & \cdots & \lambda_2^{N_\sigma-2}
	\\
	\vdots & \vdots & \vdots & \ddots &\vdots
	\\
	1 & \lambda_{N_\sigma-1} & \lambda_{N_\sigma-1}^2 & \cdots & \lambda_{N_\sigma-1}^{N_\sigma-2}
\end{pmatrix}
\begin{pmatrix}
	F_1
	\\
	F_2
	\\
	\vdots
	\\
	F_{N_\sigma -1}
\end{pmatrix}
=
\begin{pmatrix}
	R_1
	\\
	R_2
	\\
	\vdots
	\\
	R_{N_\sigma -1}
\end{pmatrix},
\end{aligned}
\ee
where
$ R_a(\lambda_a, \pi_a )$
are given by \eqref{24}. Integrals of motion  are the solutions to this equation and may  be expressed via the inverse Vandermonde matrix, explicitly,
\be
\begin{gathered}
\label{eq:F_general}
	F_\alpha =(-1)^{\alpha -1}\sum_{a=1}^{N_\sigma-1}R_a\ \frac{ A_{N_\sigma -\alpha-1}^{\ne a}}{\prod\limits_{\substack{b=1\\b\ne a}}^{N_\sigma-1}(\lambda_b-\lambda_a)}, \quad \alpha =1,...,N_\sigma -2,\\
	F_{N_\sigma-1}=\sum_{a=1}^{N_\sigma-1}\frac{R_a}{\prod\limits_{\substack{b=1\\b\ne a}}^{N_\sigma-1}(\lambda_a-\lambda_b)},
\end{gathered}
\ee
where
\be
A_\alpha^{\ne a}\equiv\sum\limits_{\substack{1\le k_1 < ... < k_{\alpha}\\k_1,...,k_{\alpha}\ne a}}^{N_\sigma-1}\lambda_{k_1}\ ...\ \lambda_{k_{\alpha}}.
\ee
In the following subsection we will first derive the explicit forms of these first integrals in the initial $x_a,\ \varphi_i$ coordinates in seven, nine and eleven dimensions \cite{Demirchian:2017uvo} and then generalize these results for arbitrary dimensions.

\subsubsection{Cases of 7, 9 and 11 dimensions}
For the simplest case of $N=3$, corresponding to seven-dimensional MP black hole, we have two integrals of motion given by \eqref{eq:F_general}:
\be
F_1=\casimir=\frac{\lambda_1R_2- \lambda_2R_1}{\lambda_1-\lambda_2},\qquad F_2=\frac{R_1-R_2}{\lambda_1-\lambda_2}
\ee
Using the expression 
\be
\label{eq:pi}
\pi_a=-\frac{1}{2}\sum_{b=1}^{N-1}\frac{p_b}{x_b}\frac{\prod\limits_{\substack{i=1\\{i\ne a}}}^{N}(m_b-\lambda_i)}{\prod\limits_{\substack{i=1\\{i\ne b}}}^{N}(m_b-m_i)}
=-\frac{1}{2}\sum_{b=1}^{N-1}\frac{p_b x_b}{m_b-\lambda_a}
\ee
one can explicitly calculate $F_2=F_{N-1}$, which is valid for any $N$ \footnote{Hereinafter, we ignore an additional constant term and  an overall constant factor which might arise in the expressions for the first integrals.}
\be
\label{eq:F_last}
F_{N-1}=\left(\sum_{a=1}^{N-1}p_a x_a\right)^2-\sum_{a=1}^{N-1}p_a^2 m_a-\sum_{i=1}^{N}\frac{m_ip_{\varphi_i}^2}{x_i^2}+g_0\sum_{i=1}^{N}x_i^2\; .
\ee


In the case of $N=4$ we will obtain all three integrals of motion from the equation \eqref{eq:F_general}:
\begin{equation}
\label{eq:nu0nu1nu2}
\begin{aligned}
F_1&=\casimir=\frac{1}{D}\bigg(R_1\lambda_2\lambda_3(\lambda_3-\lambda_2)-R_2\lambda_1\lambda_3(\lambda_3-\lambda_1)+R_3\lambda_1\lambda_2(\lambda_2-\lambda_1)\bigg),\\
F_2&=\frac{1}{D}\bigg(R_1(\lambda_2^2-\lambda_3^2)+R_2(\lambda_3^2-\lambda_1^2)+R_3(\lambda_1^2-\lambda_2^2)\bigg),\\
F_3&=\frac{1}{D}\bigg(R_1(\lambda_3-\lambda_2)+R_2(\lambda_1-\lambda_3)+R_3(\lambda_2-\lambda_1)\bigg),
\end{aligned}
\end{equation}
where \[D=\lambda_2\lambda_3(\lambda_3-\lambda_2)-\lambda_1\lambda_3(\lambda_3-\lambda_1)+\lambda_1\lambda_2(\lambda_2-\lambda_1).\]
Constants of motion $F_1$ and $F_3$ in Cartesian coordinates are given by \eqref{L} and \eqref{eq:F_last} respectively.	The second integral of motion can be derived by directly transforming the  second equation in \eqref{eq:nu0nu1nu2}. 
Using \eqref{eq:pi} we derive the expression for  $F_2=F_{N-2}$ which is valid for arbitrary $N$
\be
\label{eq:N-2}
\begin{multlined}
	F_{N-2}=\sum_{a,b=1}^{N-1}p_a x_a p_b x_b\sum_{\substack{k=1\\k\ne a,b}}^{N}m_k-\sum_{a=1}^{N-1}(p_a x_a)^2m_a+\sum_{a=1}^{N-1}p_a^2(m_a^2-f_1m_a)
	\qquad\qquad\qquad\\
	+\sum_{i=1}^{N}\frac{p_{\varphi_i}^2}{x_i^2}(m_i^2-f_1m_i)+g_0\sum_{\substack{i,j\\i\ne j}}^{N}m_i x_j^2,
\end{multlined}
\ee
where
\be
\label{eq:f1f2}
f_1(x_i,m_j)\equiv\sum_{i}^{N}(-{x_i}^2+m_i).
\ee


If $N=5$ we have four integrals of motion, three of which are given by \eqref{L}, \eqref{eq:F_last} and \eqref{eq:N-2}. The missing one is $F_2$ which, as in the previous cases, is given by \eqref{eq:F_general} in ellipsoidal coordinates
\be
F_2=\sum_{i=1}^{4}R_i
\frac{\sum\limits_{j=1}^{3}(-1)^{j-1}\lambda_i^{3-j}f_{j-1}}
{\prod\limits_{\substack{k=1\\k\ne i}}^{4}(\lambda_i-\lambda_k)},
\ee
where $f_0\equiv1$, $f_1$ is given by \eqref{eq:f1f2} and
\be
f_2(x_i,m_j)\equiv\sum_{\substack{i, j\\i \ne j}}^{N}m_i(-x_j^2+\frac{m_j}{2}).
\ee
Using \eqref{eq:pi} we can represent $F_2=F_{N-3}$ in Cartesian coordinates
and generalize it to higher dimensions
\be
\begin{aligned}
	\label{eq:N-3}
	F_{N-3}=&\sum_{j,k}^{N-1}p_jx_jp_kx_k M_2^{\ne k,j}-\sum_{j=1}^{N-1}(p_jx_j)^2(M_1^{\ne j}m_j-m_j^2)\\
	&-\sum_{j=1}^{N-1}p_j^2(m_jf_2-m_j^2f_1+m_j^3)-\sum_{j=1}^{N}\frac{p_{\varphi_j}^2}{x_j^2}(m_jf_2-m_j^2f_1+m_j^3)+g_0\sum_{i=1}^{N}x_i^2 M_2^{\ne i}
\end{aligned}
\ee
Here we use the following notations
\be
M_1^{\ne j}\equiv\sum_{\substack{k=1\\k\ne j}}^{N}m_k,\qquad M_2^{\ne j_1,..,j_a}\equiv\sum_{\substack{k_1,k_2=1\\k_1,k_2\ne j_1,..,j_a}}^{N}m_{k_1}m_{k_2}.
\ee

\subsubsection{Generalization to arbitrary dimensions}

After tedious transformations  on can  rewrite \eqref{eq:F_general} in $x_a, \varphi_i$  coordinates,
\be
\label{eq:integrals}
F_{a}=(-1)^a\sum_{b,c=1}^{N_\sigma-1}K^{bc}_{(a)}(x)p_bp_c- \sum_{i,j=1}^{N}L^{ij}_{(a)}p_{\varphi_i}p_{\varphi_j}+(-1)^{a-1}A_{N_\sigma-a}m_0^2r_H^2,
\ee
where

\begin{align}
&\begin{multlined}[t]
	K^{bc}_{(a)}=
	\left(\sum_{\alpha=0}^{N_\sigma-a-1}(-1)^{N_\sigma+\alpha-a}A_{\alpha}m_b^{N_\sigma-\alpha-a}
	+x^2_b\sum _{\alpha=1}^{N_\sigma-a-1}(-1)^{\alpha}M_{N_\sigma-\alpha-a-1}^{\ne b}m_b^{\alpha }
	\right)\delta^{bc}
	\quad\\
	+M_{N_\sigma-a-1}^{\ne b,c}x_bx_c
\end{multlined}
\\
&\begin{multlined}[t]
	L^{ij}_{(a)}=\left((1-\delta_a^1)\sum_{\alpha=1}^{N_\sigma-a}(-1)^{N_\sigma+\alpha}A_{\alpha-1}m_i^{N_\sigma-a-\alpha+1}-\delta_a^1A_{N_\sigma-1}\right)\frac{\delta^{ij}}{x^2_i}
	\qquad\qquad\qquad\\
	+(-1)^{a-1}A_{N_\sigma-a}\frac{\sqrt{m_i-1}}{m_i}\frac{\sqrt{m_j-1}}{m_j}
\end{multlined}
\end{align}

with
\be
\begin{aligned}
	\label{eq:f}
	A_a(x_i,m_j)&\equiv \sum\limits_{1\le k_1 < ... < k_{a}}^{N_\sigma-1}\lambda_{k_1}\ ... \ \lambda_{k_{a}}\\
	&=-\sum_{i=1}^{N_\sigma}x^2_iM^{\ne i}_{a-1}+
	\sum\limits_{\substack{1\le k_1 < ... < k_{a}}}^{N_\sigma}m_{k_1}\ ... \ m_{k_{a}}
	, \qquad a=1,\ldots,N_\sigma-1,
\end{aligned}
\ee
and
\be
M^{\ne a_1, ..., a_j}_i\equiv \sum\limits_{\substack{1\le k_1 < ... < k_{i}\\k_1,...,k_i\ne a_1, ..., a_j}}^{N_\sigma}m_{k_1}\ ... \ m_{k_{i}} , \qquad j=0,\ldots,N_\sigma-1,\quad i=1,\ldots,N_\sigma-j.
\ee
It is also assumed that
\be
A_0\equiv 1, \qquad M^{\ne a_1, ..., a_j}_0\equiv 1.
\ee
One can check that in odd dimensions in the special cases of $F_{N-1},\ F_{N-2}$ and $F_{N-3}$, the above reduce to the corresponding integrals of motion \cite{Demirchian:2017uvo} given by \eqref{eq:F_last}, \eqref{eq:N-2} and \eqref{eq:N-3}.
One can also check,  that  simply requiring the rotation parameters to be equal in these expressions, one does not recover all the integrals of the special case of $a_i=a, \forall i$  NHEMP. In such special cases all of the first integrals of the spherical mechanics of generic (non-equal $a_i$) case  transform into the Hamiltonian of the spherical mechanics of the equal $a_i$ case. So, to obtaining the Liouville integrals in the isotropic case we need to develop more sophisticated contraction procedure.

We also note that  the above expressions for the constants of motion  were found  in  the ellipsoidal coordinates introduced  for the special case of {\sl non-equal} rotational parameters $a_i$.
However, we then written  them   in the initial coordinates, they hold   for generic nonzero values of the rotation parameters $a_i$.
We will analyze the special cases where some of the $a_i$ or $m_i$ are equal in \sect{\ref{Special-cases-Sec}} and when one of them is vanishing in \sect{\ref{EVH-Sec}}.

\subsubsection{Killing tensors}
In previous subsection  we presented the constants of motion in the form  demonstrating their  explicit dependence on the momenta $p_a, p_{\varphi_i}$.
To represent \eqref{eq:integrals} through the respective $N_\sigma$ second rank Killing tensors, one can replace the last term proportional to $m_0^2$ from  the mass-shell equation \eqref{mass-shell-1},  \eqref{eq:mass_shell}.  Note also that the $F_a,\  a=1,\ldots.N_\sigma-1$, provides $N_\sigma-1$ one constants of motion. We can then add $F_{N_\sigma}$ to this collection, which is proportional to the mass with the corresponding second rank killing tensor being the inverse metric, i.e.
\be	
F_{N_\sigma}=(-1)^{a-1}\left(r^2_H\sum_{A,B=1}^{2N+1+\sigma}g^{AB}p_A p_B-\left(\sum_{i=1}^N \frac{\sqrt{m_i-1}p_{\varphi_i}}{m_i}\right)^2\right),
\ee
where we assumed	$M_{-1}^{\ne b,c}=0$.

To get the expression for Killing tensors, we should simply replace the momenta by the respective vector fields, $p_A\to \frac{\partial}{\partial x^A}$.
That is,  in the coordinates $(x_a, \varphi_a)$  where the constants of motion  \eqref{eq:integrals} are written,
one should replace
$$
p_a \to \frac{\partial}{\partial x^a}, \quad p_{\varphi_i} \to \frac{\partial}{\partial \varphi^i},\quad p_r\to \frac{\partial}{\partial r},\quad p_0\to\frac{\partial}{\partial \tau}.
$$
In ellipsoidal coordinates the  above presented  $N_\sigma-2$ Killing tensors read
\be
\begin{multlined}
	K_{a}=\sum_\alpha A_\alpha^{\neq a}\, h^\alpha\, \left(\partial_{\lambda_\alpha}\right)^2
	+\sum_I\sum_\alpha \frac{A_\alpha^{\neq a}\,{\prod_{J\neq I}\,} (m_J-m_I)}{m_I(m_I-\lambda_a){\prod_{b}\,}'(\lambda_b-\lambda_a)}\,\left(\partial_{\varphi_I}\right)^2
	\qquad\qquad\\
	+\frac{A^{a}}{A(\lambda)}\left(-\frac{1}{r^2}\left(\partial_\tau\right)^2+r^2 \left(\partial_r\right)^2\right).
\end{multlined}
\ee
Thus,  we have $N+1$ mutually commuting Killing vectors $\partial/\partial\varphi_i, \partial /\partial_\tau$
and $N_\sigma$ Killing tensors, summing up to $d=N_\sigma+N+1$ and hence the system is integrable. One may check that our expressions for the Killing tensors match with those appeared in \cite{Hidden-symmetry-NHEK, Kolar:2017vjl}  after taking the near-horizon limit. We note that the two extra Killing vectors of the $SL(2,\mathbb{R})$ part of the isometry which appear in the near horizon limit and in the coordinates of
\eqref{m2} take the form,
\be
r\frac{\partial}{\partial r}-\tau\frac{\partial }{\partial \tau},\qquad (\tau^2+\frac{1}{r^2})\frac{\partial }{\partial \tau}-2\tau r\frac{\partial}{\partial r}-\frac{2}{r}\sum_{i=1}^N \frac{\partial}{\partial\varphi_i},
\ee
do not yield new independent constants of motion.

\subsection{The fully isotropic, equal rotation parameters}\label{fully-isotropic-sec}

When all of the rotational parameters coincide, the Hamiltonian of probe particle  reduces to the system on sphere and admits separation of variables in spherical coordinates \cite{GNS-11,GNS-12}.
It can be checked that in this case, the Hamiltonian of the reduced mechanics derived from \eqref{L} transforms into the corresponding mechanics with equal parameters derived in \cite{GNS-11,GNS-12}
for both odd and even dimensional cases.
Notice,  that in this limit the difference between even and odd cases becomes visible:
\begin{itemize}
	\item
	In the odd case, $\sigma=0$, isotropic limit corresponds to the choice $m_i=N$ , $i=1,\ldots, N$.
	As a result, the angular Hamiltonian \eqref{L} which we will denote it by ${\cal I}_N$ takes the form
	\be
	\mathcal{I}_N=\sum\limits_{a,b=1}^{N-1}(N\delta_{ab}-{x_a x_b})p_{a}p_{b}+N\sum\limits_{i=1}^{N}\frac{p_{\varphi_i}^2}{x_i^2},\qquad \sum_{i=1}^N x_i^2=N.
	\label{oddIsoHam}\ee
	For the fixed $p_{\varphi_i}$ configuration space of this system is $(N-1)$-dimensional sphere, and   the  Hamiltonian defines specific generalization
	of the Higgs oscillator, which is also known as a Rossochatius system \cite{Rosochatius}.
	
	\item In the even case, $\sigma=1$, one has $m_i=2N$ when $i=1,\ldots, N$ and $m_{N+1}=1$, i.e. we can't choose all parameters $m_I$ be equal.
	As a result, the angular Hamiltonian \eqref{L} reads
	\be
	\mathcal{I}_N=\sum\limits_{i,j=1}^{N}(\eta^2\delta_{ij}-{x_ix_j})p_{i}p_{j}
	+\sum\limits_{i=1}^{N}\frac{\eta^2p_{\varphi_i}^2}{x_i^2}
	+\omega\sum_{i=1}^Nx_i^2,
	\label{evenIsoHam}\ee
	where
	\be
	\eta^2=4N^2- (2N-1)\sum_{i=1}^Nx_i^2,\qquad
	\omega=\left(1-\frac{1}{2N}\right)^2\sum_{i,j=1}^{N}p_{\varphi_i}p_{\varphi_j}-m_0^2\,(2N-1).
	\ee
	In the case of even dimension configuration space fails to be sphere (even with  fixed $p_{\varphi_i}$).
\end{itemize}
What is important is that both systems admit separation of variables in spherical coordinates.
Namely, by recursively introducing spherical coordinates
\be
x_{N_\sigma}=\sqrt{N_\sigma}\cos\theta_{N_\sigma-1}, \qquad x_{a}=\sqrt{N_\sigma}{\tilde x}_a\sin\theta_{N_\sigma-1}, \qquad \sum_{a=1}^{N_\sigma-1}{\tilde x}_a^2=1,
\label{sphin}\ee
we get the following recurrent formulae for the constants  of motion
\bea
\label{eq:iso_integrals_odd}
&\sigma=0:&\mathcal{I}_{odd}=F_{N-1},\quad
F_a=p_{\theta_{a}}^2+\frac{p_{\varphi_{a+1}}^2}{\cos^2\theta_{a}}+\frac{F_{a-1}}{\sin^2\theta_{a}},\quad
F_0=p_{\varphi_1}^2\label{oddconstseq}\\
&\sigma=1 : & \mathcal{I}_{even}=2 N p^2_{\theta_{N}}+\nu\sin^2\theta_{N} + \left(2 N\cot^2\theta_{N}+1\right)F_{N-1},
\label{evenconsteq}
\eea
It is clear, that  $F_1,\ldots, F_{N_\sigma -1}$ define complete set of Liouville   constants of motion and  the $\sigma=1$ system contains $\sigma=0$ as a subsystem.
Moreover,  the  Rosochatius system (angular Hamiltonian for $\sigma=0$ case with fixed $p_{\varphi_i}$) is superintegrable: it has $N-2$ additional functionally independent constants of motion  defined by the expression
\begin{align}
\begin{multlined}[t]
\label{eq:hidden_integrals}
I_{a,a-1}=\left(p_{\theta_{a-2}}\sin\theta_{a-2}\cot\theta_{a-1}-p_{\theta_{a-1}}\cos\theta_{a-2}\right)^2+
\qquad\qquad\qquad\\
\left(p_{\varphi_{a-1}}\frac{\cot\theta_{a-1}}{\cos\theta_{a-2}}+p_{\varphi_a}\cos\theta_{a-2}\tan\theta_{a-1}\right)^2.
\end{multlined}
\end{align}
When $p_{\varphi_i}$ are not fixed, the system is $(N_\sigma -1+ N)$-dimensional one.  In that case, from its action-angle formulation \cite{GNS-11,GNS-12} one can observe, that
it remains maximally superintegrable for $\sigma=0$, i.e. possesses $4N-3$ constants of motion:
Besides   $2N-3$ constants of motion  given by \eqref{oddconstseq} and \eqref{eq:hidden_integrals}, and
the $N$ commuting integrals  $p_{\varphi_i}$ (associated with axial Killing vectors), there are  $N$ additional  constants of motion with quadratic term mixing $p_{\theta_a}$ and $p_{\varphi_i}$; i.e. $N$ second rank Killing tensors in $\partial_{\theta_a}\partial_{\varphi_i}$ direction.
When $\sigma=1$, the system is $2N$-dimensional, and has  $4N-2$ integrals, i.e., as lacks one integral from being maximal superintegrable.

From these constant of motion one can readily read the associated Killing vectors and second rank Killing tensors.
Hence, isotropic system has $N+1$ mutually commuting Killing vectors and  $d-3=2N+\sigma-2$ Killing tensors, and an additional $N$ non-commuting second rank Killing tensors.


For more detailed  analysis of the isotropic case see \cite{GNS-11,GNS-12}. Here we present it mainly to set the conventions we use in the study of ``intermediate case", when only some of the rotation parameters are equal to each other.

%% file: Integrability_of_MP_2.tex
\addtocontents{toc}{\protect\newpage}  
\section[Integrability of geodesics in general non-vanishing and vanishing NHEMP geometry]
		{Integrability of geodesics 
		\\ in general non-vanishing and vanishing NHEMP geometry}
\label{sec:Integrability_of_MP_2}

In the previous chapter we studied two special cases of the integrability of Hamilton-Jacobi equation in the background of Non-Vanishing Near Horizon Extremal Myers-Perry black hole: when there are no equal rotation parameters and when all of the rotation parameters are equal. In this chapter we are going to generalize the previous results and study the case when some of the rotational parameters are equal and others are not. Furthermore, we are going to study the integrability of geodesics in the background of vanishing NHEMP geometry (when one of the rotational parameters is 0).

\subsection{Partially isotropic cases}\label{Special-cases-Sec}

When some of the $a_i\neq 0$'s are equal the geometry \eqref{NHEMP-metric-rt-coord} exhibits a bigger isometry group than $SL(2,\mathbb{R})\times U(1)^N$;  depending on the number of equal $a_i$'s the $U(1)^N$ part is enhanced to a rank $N$ subgroup of $U(N)$. This larger isometry group brings larger number of Killing vectors and tensors and one hence expects the particle dynamics for these cases to become  a superintegrable system. This is what we will explore in this section and construct the corresponding conserved charges.

\subsubsection{Partially isotropic case in odd dimension}

Let's start with the simpler odd dimensional system, $\sigma=0$, with $p=N-l$  nonequal rotation parameters and  $l$ equal ones:
\be
m_1\ne m_2\ne\ldots\ne m_p\ne m_{p+1},\quad  m_{p+1}=m_{p+2}= \ldots = m_{N}\equiv\kappa.
\ee
Starting from the metric \eqref{m2} we will construct the Hamiltonian for the reduced mechanics by introducing spherical and ellipsoidal coordinates.
Spherical coordinates $\{y,\ \theta_i\},\ i=1\ldots l-1$ will be introduced for the $l$ latitudinal coordinates $x_{p+1},\ldots, x_{N}$ corresponding to the equal rotational parameters
\be
\label{eq:shpere_intremediate_special}
\begin{gathered}
x_{p+1}=y\prod_{i=1}^{l-1}\sin{\theta_{i}},\qquad x_{p+a}=y	\cos{\theta_{a-1}}\prod_{i=a}^{l-1}\sin{\theta_{i}},\quad x_{p+l}=y \cos{\theta_{l-1}},\qquad a=2,\ldots, l-1.
\end{gathered}
\ee
Hence,
\be
\sum_{a=1}^{l}\frac{x^2_{p+a}}{m_{p+a}}=\frac{y^2}{\kappa}, \qquad
\sum_{a=1}^{l}(dx_{p+a})^2=(dy)^2+y^2\ d\Omega_{l-1},
\ee
with $d\Omega_{l-1}$ being the metric on $(l-1)$-dimensional sphere: $d \Omega_{l-1}=d\theta_{l-1}^2+\sin^2\theta_{l-1} d \Omega_{l-2}$.

Performing the coordinate transformation \eqref{eq:shpere_intremediate_special} in \eqref{NHEMP-metric-rt-coord}, it is seen that the radial coordinate $y$ of the spherical subsystem behaves very much like the other latitudinal coordinates of non-equal rotational parameters. Therefore, we will treat $y$ and $x_1\ldots x_p$ in the same way:
\begin{align}
\begin{multlined}
	y_a=(x_1,\ldots,x_p, y),\qquad \widetilde{m}_a=(m_1,\ldots,m_p, m_{p+1})\;:
	\qquad\qquad\qquad\qquad \\
	\sum_{a=1}^{p+1}\frac{1}{\widetilde{m}_a}=1,\quad  \sum_{a=1}^{p}\frac{y_a^2}{\widetilde{m}_a}+{\frac{y^2}{\widetilde{m}_{p+1}}}=1,
\end{multlined}
\end{align}
in terms of which the metric takes the form
\be
\label{eq:metric_intermediate_special}
\frac{ds^2}{r^2_H}=A(y)\left(-r^2d\tau^2+\frac{dr^2}{r^2}\right)+
dy_{p+1}^2+y_{p+1}^2d\Omega_{l-1}+\sum_{a=1}^{p}
(dy_a)^2 +
\sum_{i,j=1}^N\tilde{\gamma}_{ij}x_i(y) x_j(y) D\varphi^iD\varphi^j,
\ee
with
\be
A(y) ={\frac{\sum_{a=1}^{p+1} y^2_a/\widetilde{m}^2_a}{4\sum_{a<b}^{p+1}\frac{1}{\widetilde{m}_a}\frac{1}{\widetilde{m}_b}}}, \qquad \sum_{a=1}^{p+1}\frac{y_a^2}{\widetilde{m}_a}=1.
\label{metspec}\ee

Hamiltonian of the corresponding spherical mechanics then reads
\be
\begin{gathered}
\label{eq:red_mech_inter_special}
	\mathcal{I}=A\left[\sum_{a,b=1}^{p}h^{ab}p_a p_b+\sum_{a=1}^{p+1}\frac{g_a^2}{y_a^2}+g_0\right],
	\\
	{\rm with}\quad
	g_a^2=(p^2_{\varphi_1},\ldots, p^2_{\varphi_p}, \mathcal{I}_{p+1}),\quad  h^{ab}=\delta^{ab}-\frac{1}{\sum\limits_{a=1}^{p+1} y^2_a/\widetilde{m}^2_a}\frac{y_a}{\widetilde{m}_a}\frac{y_b}{\widetilde{m}_b}.
\end{gathered}
\ee
and $\mathcal{I}_{p+1}$ defined as  by \eqref{oddIsoHam} in $(p+1)$-dimensional space.
The above describes a  lower-dimensional version of \eqref{L}, where all rotational parameters are nonequal and we can analyze it as we did for the general case in the previous chapter. That is,
we introduce on the $(p+1)$-dimensional ellipsoidal coordinates
\be
\label{eq:ellips_intremediate_special}
y_a^2=\frac{\prod_{b=1}^{p+1}\left(\widetilde{m}_a-\lambda_b\right)}{\prod^{p+1}_{b=1; b\neq a}\left(\widetilde{m}_a-\widetilde{m}_b\right)},
\ee
and take $\lambda_{p+1}=0$ for  resolving the constraint \eqref{metspec} given by the second expression. The rest of the analysis goes through as  in \cite{non-equal-general}  and as in \sect{\ref{Particle-dynamics-Sec}}.

The partially isotropic case discussed here, as we see, interpolates between the generic case of \sect{\ref{Particle-dynamics-Sec}} ($p=N-1$) and the fully isotropic case ($p=0$) of \sect{\ref{fully-isotropic-sec}}: it decouples to the  Hamiltonians of type \eqref{eq:conf_ham_odd} and  \eqref{oddIsoHam}.
The case $l=1$ corresponds to  the system with non-equal parameters, and  the spherical subsystem is trivial ($\mathcal{I}_{p+1}=p_{\varphi_{p+1}}^2$).
For  $l\geq 2$ the $(l-1)$-dimensional  spherical subsystem is not trivial anymore and has $2(l-1)-1$  constants of motion.
Thus  the reduced $(N-1)$-dimensional angular system has $p+2l-3 =N-1+l-2$ constants of motion, i.e. the  number of extra constants of motion  compared to the generic case is $l-2$, with $l>2$.
It becomes maximally superintegrable  only for $l=N$, i.e when  all rotational parameters are equal.

This discussion can be easily extended to the case of even dimensions ($\sigma=1$). Here we will have an additional latitudinal coordinate ($p+l=N+1$) and a rotational parameter with a fixed value ($m_{N+1}=1$). One should note that $m_{N+1}$ cannot be equal to any other rotational parameter,  so it is one of the $p$ non-equal parameters.  In the limiting case when $l=1$ and all rotational parameters are different and we have an integrable system with $p=N$ configuration space degrees of freedom, as expected. Since $m_{N+1}$ cannot be equal to the others, $p$ cannot be equal to $0$ and the even dimensional system cannot be maximally superintegrable. In the limit when all rotational parameters are equal except $m_{N+1}$ ($p=1$), the system will lack one integral of motion to be maximally superintegrable.

\subsubsection{General case}
\label{sec:gc}

Having discussed the some equal $m_i$'s but the rest nonequal case, we now turn to the  most general case when there are  $s$ sets (blocks) of equal rotation parameters each containing $l_i$ members. As before we  assume that there are $p$ rotation parameters which are not equal to the others, so that $p+\sum_{i=1}^s l_i=N_\sigma$. Note that in our conventions $l_i\geq 2$. We introduce an upper index which, written on a parameter or a function, denotes the number of the block under consideration. So, for example $m_a^{(i)}$ will denote all the equal rotational parameters in the $i$-th set of rotation parameters and $x_a^{(i)}$ will denote their corresponding latitudinal coordinates and
\be
\begin{gathered}
\{m_a^{(i)}\}=m_{p+l_1+\ldots +l_{i-1}+a}\equiv\kappa^{(i)}\qquad
i=1,\ldots,s,\qquad a=1,\ldots,l_i.
\end{gathered}
\ee
The list of all rotational parameters can be written as
\be
\begin{gathered}
\{m_\alpha\}=m_1,\ m_2,\ldots,\ m_p,\ \{m_a^{(1)}\},\ \{m_a^{(2)}\},\ldots,\ \{m_a^{(s)}\},\qquad \alpha=1,\ldots,N\\
m_1\ne m_2\ne\ldots\ne m_p,\qquad \{m_a^{(i)}\}=\kappa^{(i)}\quad
\text{with}\qquad \kappa^{(i)}\ne\kappa^{(j)}, \quad p+l_1+\ldots +l_{s}=N.	
\end{gathered}
\ee

Let us start with the odd  ($\sigma=0)$ case and the metric \eqref{m2}. We can construct the Hamiltonian for the reduced mechanics by introducing spherical and ellipsoidal coordinates. Different spherical coordinates will be introduced separately for each set of latitudinal coordinates corresponding to different sets of equal rotational parameters.
\be
\label{eq:shpere_intremediate}
x_1^{(i)}=r_i\prod_{\alpha=1}^{l_i-1}\sin{\theta_{\alpha}^{(i)}}\qquad	x_k^{(i)}=r_i\cos{\theta_{k-1}^{(i)}}\prod_{a=k}^{l_i-1}\sin{\theta_{a}^{(i)}}\qquad
x_{l_i}^{(i)}=r_i\cos{\theta_{l_i-1}^{(i)}},\qquad
k=2,\ldots, l_i-1
\ee
One should note that these spherical coordinates satisfy the  relations
\be
\sum_{a=1}^{l_i}(x_a^{(i)})^2=r_i^2 \quad
\text{and} \quad
\sum_{a=1}^{l_i}(\diff x_a^{(i)})^2={\diff r_i}^2+r_i^2d \Omega^{(i)}_{l_i-1},
\ee
where $d\Omega^{(i)}_{n}=(\diff {\theta^{(i)}_{n}})^2+\sin^2\theta^{(i)}_{n}d\Omega^{(i)}_{n-1}$ denotes
the metric on unit $n$-dimensional sphere.
For the rest of the latitudinal coordinates $x_1\ldots x_p$ corresponding to non-equal rotational parameters and the radial coordinates $r_i$ of  isotropic subsystems
we introduce the notation
\be
\label{eq:ellips_intremediate}
\{y_a\}=\{x_1,\ldots,x_p;\ r_1,\ldots, r_s\},\qquad
\{\widetilde{m}_a\}=\{m_1,\ldots,m_p;\ \kappa^{(1)},\ldots, \kappa^{(s)}\},
\ee
In this notation the  the metric \eqref{m2} can be rewritten as
\be\label{eq:metric_intermediate}
\frac{ds^2}{r^2_H}=A(y)\left(-r^2d\tau^2+\frac{dr^2}{r^2}\right)+\sum_{a=1}^{p+s} {dy_a}^2+\sum_{b=1}^sy_{p+b}^2d\Omega^{(b)}_{l_b-1}+
\sum_{i,j=1}^N\tilde{\gamma}_{ij}x_i(y) x_j(y) D\varphi^iD\varphi^j,
\ee
where   $\tilde\gamma_{ij}$ , $A(y)$ are defined as in \eqref{gamma} and \eqref{metspec} respectively.
Therefore,  the Hamiltonian of the corresponding angular mechanics reads
\be
\begin{gathered}
	\label{eq:red_mech_inter}
	\mathcal{I}=A\left[\sum_{a,b=1}^{p+s-1}h^{ab}\pi_a\pi_b+\sum_{a=1}^{p+s}\frac{g_a^2}{y_a^2}+g_0\right]
	\\
	\{g_a^2\}=\{p^2_{\varphi_1},\ldots, p^2_{\varphi_p};\ \mathcal{I}^{(1)},\ldots,\mathcal{I}^{(s)}\}, \qquad \mathcal{I}^{(a)}=F^{(a)}_{l_a-1},	
\end{gathered}
\ee
where $\mathcal{I}^{(a)}$ are the spherical subsystems resulting from the $s$ sets of equal rotation parameters, $h^{ab}$ is defined by \eqref{eq:red_mech_inter_special}, and
\be
\begin{gathered}
F_{d}^{(a)}=p_{\theta_{d}^{(a)}}^2+\frac{(g_{d+1}^{(a)})^2}{\cos^2\theta_{d}^{(a)}}+\frac{F_{d-1}^{(a)}}{\sin^2\theta_{d}^{(a)}}, \qquad
F_{0}^{(a)}=(g_{1}^{(a)})^2, \qquad g_{d}^{(a)}=p_{\varphi_{p+l_1+\ldots+l_{a-1}+d}}\\
\{\pi_a,\lambda_b\}=\delta_{ab},\qquad
\{p_{\varphi_i},\varphi_j\}=\delta_{ij}\qquad
\{p_{\theta_{b}^{(a)}},\theta_{d}^{(c)}\}=\delta_{ac}\delta_{bd},
\end{gathered}
\ee
Hence, the reduced spherical mechanics \eqref{eq:red_mech_inter} has the exact form of \eqref{eq:conf_ham_odd} (with appropriate constants) whose integrability has already been discussed.
All discussions from the previous subsection can be easily extended to this case, e.g.  separation of variables
may be achieved in the ellipsoidal coordinates
\be
y_a^2=\frac{\prod_{b=1}^{{p+s}}\left(\widetilde{m}_a-\lambda_b\right)}{\prod^{{p+s}}_{b=1; b\neq a}\left(\widetilde{m}_a-\widetilde{m}_b\right)},
\ee
and place $\lambda_{p+s}=0$ for resolving the constraint on  latitudinal coordinates \eqref{eq:restr_on_x}, which  now takes the form
$\sum_{a=1}^{p+s}\frac{y_a^2}{\widetilde{m}_a}=1$.

So, we separated the variables for the $(N-1)$-dimensional angular mechanics describing the geodesics in the near-horizon limit of $(2N+1+\sigma)$-dimensional Myers-Perry black hole  in arbitrary dimension
with arbitrary non-zero values of rotational parameters.
The number of constants of motion in this system  can be easily counted: it is equal to $d+N_\sigma-p-2s$. The generic case of nonequal $m_i$ is recovered by $s=0, p=N_\sigma$ and the fully isotropic case as $s=1, p=0$.
In a similar manner one can construct associated Killing tensors.

\subsubsection[Contraction from fully non-isotropic to isotropic NHEMP]
				{Contraction from fully non-isotropic \\ to isotropic NHEMP  }\label{sec:contrraction}

Having the two corner cases discussed (fully non-isotropic and isotropic) an interesting question arises. What kind of approximation would transform the first integrals of fully non-isotropic NHMEP to the first integrals of isotropic NHEMP? It is straightforward to check that simply taking all rotation parameters to be equal just transforms all the first integrals of fully non-isotropic NHMEP to the Hamiltonian of the spherical mechanics of isotropic NHEMP (with an overall constant factor and a constant term). So if $m_i=N$
\be
F_a=
C_a\left(\sum\limits_{b,c=1}^{N-1}(\delta_{bc}-{x_b x_c})p_{b}p_{c}+\sum_{k=1}^N \frac{p^2_{\varphi_k}}{x_k^2}\right)+C^\prime_a
\ee
where $C_a$ and $C^\prime_a$ are constants. To find the desired approximation, we will work with rotation parameters which have little variations from their isotropic value $N$ ($\epsilon_i\ll N$), \[m_i=N+\epsilon_i.\] In such a limit, the Hamiltonian of the non-isotropic mechanics can be extended in powers of $\epsilon_i$, keeping the first order term only
\be
\begin{aligned}
	\label{eq:F_1}
	F_1=N^{N-3}\left[N\casimir_{iso}+N^2g_0
	-\sum_{i=1}^{N}\epsilon_ix_i^2 \left[\sum_{a}^{N-1}p_a^2+\sum_{k}^{N}\frac{p_{\varphi_k}^2}{x_k^2}+g_0\right]+2\sum_{a,b}^{N-1}\epsilon_a p_ax_ap_bx_b\right]
\end{aligned}
\ee
where
\be
\casimir_{iso}=\sum\limits_{a,b=1}^{N-1}(N\delta_{ab}-{x_a x_b})p_{a}p_{b}+N\sum\limits_{i=1}^{N}\frac{p_{\varphi_i}^2}{x_i^2}
\ee
is the isotropic Hamiltonian. We should note that the linear term of $F_1$ still corresponds with the isotropic Hamiltonian $\casimir_{iso}$ but the relation $\sum x_i^2=N$ doesn't hold anymore.

Now, if we find some linear combination $P(F_a)$ of first integrals of non-isotropic mechanics such that the free term of the expansion around $m_i=N$ vanishes, we can write
\begin{equation}
\label{eq:iso_integ}
\begin{aligned}
\{P(F_a),F_1\}&=0=\left\{\sum_{i=1}^{N}\epsilon_iP_i(p_j,x_j)\ ,\ \casimir_{iso}+\sum_{i=1}^{N}\epsilon_i(...)\right\}=\sum_{i=1}^{N}\epsilon_i\left\{P_i(p_j,x_j)\ ,\ \casimir_{iso}\right\}\\
&\implies \qquad  \left\{P_i(p_j,x_j)\ ,\ \casimir_{iso}\right\}=0
\end{aligned}
\end{equation}
We see  that the first order coefficients $P_i(p_j,x_j)$ of the $P(F_a)$ linear combination are first integrals for $\casimir_{iso}$. To construct such combination whose free term vanishes we can take any of the first integrals, let's say $F_{N-1}$ and expand it.
\be
\label{eq:F_N-1}
\begin{aligned}
	F_{N-1}=(-1)^N&\left[\casimir_{iso}-\frac{g_0}{N}\sum_{i=1}^{N}\epsilon_ix_i^2+\sum_{i}^{N}\epsilon_i\frac{p_{\varphi_i}^2}{x_i^2}+\sum_{a=1}^{N-1}\epsilon_a p_a^2\right]
\end{aligned}
\ee
We see from \eqref{eq:F_1} and \eqref{eq:F_N-1} that by combining $F_1$ and $F_{N-1}$ the free term can be eliminated
\be
\label{eq:P_comb}
\begin{aligned}
	N^{-(N-3)}F_1&+(-1)^{N-1}NF_{N-1}-g_0N^2=\\
	&-\left(\sum_{a}^{N-1}p_a^2+\sum_{k}^{N}\frac{p_{\varphi_k}^2}{x_k^2}\right)\sum_{i=1}^{N}\epsilon_ix_i^2+2\sum_{a,b}^{N-1}\epsilon_a p_a x_a p_b x_b -N\left(\sum_{i}^{N}\epsilon_i\frac{p_{\varphi_i}^2}{x_i^2}+\sum_{a=1}^{N-1}\epsilon_a p_a^2\right)
\end{aligned}
\ee
Furthermore, from the expression $\sum_{i}^{N}x_i^2/m_i=1$ we can find \[x_N^2=\left(\tilde{x}_N^2+\frac{1}{N}\sum_{a}^{N-1}\epsilon_ax_a^2\right)\left(1+\frac{\epsilon_N}{N}\right),\quad \tilde{x}_N^2\equiv N-\sum_{a}^{N-1}x_a^2\]
and replace with this relation every occurrence  of $x_N$ in \eqref{eq:P_comb}. Doing this, we will end up with the same equation \eqref{eq:P_comb} with just $x_N^2$ replaced by $\tilde{x}_N^2$. So in further calculations we are free to consider equation \eqref{eq:P_comb} with a redefined $x_N$
\be
\tilde{x}_N^2 \rightarrow x_N^2=N-\sum_{a}^{N-1}x_a^2
\ee
Thus, having in mind \eqref{eq:iso_integ}, we find the first integrals of isotropic mechanics to be
\be
\begin{aligned}
	F_a^{iso}&=-x_a^2\left(\sum_{b}^{N-1}p_b^2+\sum_{k}^{N}\frac{p_{\varphi_k}^2}{x_k^2}\right)+2 p_a x_a \sum_{b}^{N-1}p_b x_b -N\left(\frac{p_{\varphi_a}^2}{x_a^2}+ p_a^2\right)\\
	F_{N}^{iso}&=-x_N^2\left(\sum_{b}^{N-1}p_b^2+\sum_{k}^{N}\frac{p_{\varphi_k}^2}{x_k^2}\right)-N\frac{p_{\varphi_N}^2}{x_N^2}
\end{aligned}
\ee
Now, we can see that the sum of all $N$ first integrals results into the casimir of isotropic mechanics
\be
\sum_{i=1}^NF_i^{iso}=-2\ \casimir_{iso}.
\ee
Thus, by definition, all $F_i^{iso}$ commute with $\sum_{i=1}^NF_i^{iso}$, but one can check that they don't commute with each other.

\subsection{Extremal vanishing horizon case}\label{EVH-Sec}
As  seen from metric \eqref{NHEMP-metric-rt-coord}, the case where one of the $a_i$'s is zero is a singular case. In fact for this case one should revisit the near-horizon limit. It has been shown that \cite{NHEVH-MP} for the odd dimensional extremal MP black holes the horizon area also vanishes and we are hence dealing with an Extremal Vanishing Horizon (EVH) black hole \cite{NHEVH-1}. The near-horizon EVH black holes have remarkable features which are not shared by generic extremal black holes; they constitute different set of geometries which should be studied separately \cite{NHEVH-three-theorems}. In particular, it has been proved that for EVH black holes the near horizon geometry include an AdS$_3$ factor (in contrast with the AdS$_2$ factor of general extremal case) \cite{NHEVH-three-theorems, Sadeghian:2017bpr}, i.e. the $d$ dimensional NHEVHMP exhibits $SO(2,2)\times U(1)^{N-1}$ isometry.
To study this case, we start by a review on black hole geometry itself.
Then, by taking the near horizon and EVH limit, we discuss the separability of
Hamilton-Jacobi equations on the NHEVH geometries.

As discussed in the special case of EVH black holes one has to revisit the standard NH theorems for extremal black holes. Here we review EVH black holes in the  family of odd dimensional MP black holes \cite{mp}:
\be\label{MP-odd}
ds^2=-d\tau^2+\frac{\mu \rho^2}{\Pi F}(d\tau+\sum_{i=1}^N a_i \mu_i^2d \phi_i)^2+\frac{\Pi F}{\Pi-\mu \rho^2} d\rho^2+\sum_{i=1}^N (\rho^2+a_i^2)(d\mu_i^2+\mu_i^2d\phi_i^2)
\ee
where
\be
F=1-\sum_i\frac{a_i^2\mu_i^2}{\rho^2+a_i^2},\quad \Pi=\prod_{i=1}^N (\rho^2+a_i^2),\qquad \sum_i \mu_i^2=1.
\ee
The extremal case happens when $\Pi-\mu \rho^2=0$ has double roots and the EVH case is when one of $a_i$ parameters, which we take to be $a_N$ is zero. That is in the EVH case $\mu=\prod_{a=1}^{N-1} a_a^2$. We note that we could have considered  a ``near-EVH'' metric where the black hole is at a non-zero but small temperature and the horizon area is also small, while the ratio of horizon area to the temperature is finite \cite{NHEVH-1, NHEVH-three-theorems}.

The horizon for the EVH case is at $\rho=0$ and hence in the NH limit, the leading contributions come from
\be\label{NHEVH-limit}
\Pi=\mu \rho^2 (1+\frac{\rho^2}{r_0^2}),\qquad F_0=1-\sum_{a=1}^{N-1} \mu_a^2,\qquad \frac{1}{r^2_0}=\sum_{b=1}^{N-1}\frac{1}{a_b^2}.
\ee
Plugging the above into the metric \eqref{MP-odd} and taking:
$$\rho=r_0\, r\,\epsilon,\quad \tau=r_0\, t/\epsilon,\quad \psi=\varphi_N/\epsilon,\quad \varphi_a=\phi_a+\tau/a_a,\quad a=1,\ldots, N-1, \qquad \epsilon\to 0,$$
we obtain the NHEVHMP metric \cite{NHEVH-MP}:
\be
\begin{gathered}
\label{NHMPEVH-metric}
	ds^2=F_0\,r^2_0\,\left[-r^2\, dt^2+\frac{dr^2}{r^2}+r^2d\psi^2\right]+ \sum_{b =1}^{N-1}a_b^2d\mu_b^2+\sum_{a,b =1}^{N-1}\gamma_{ab}d\varphi_ad\varphi_a,
	\\
	\gamma_{ab}\equiv a^2_a \mu_a^2\delta_{ab}+ a_a a_b \frac{\mu_a^2\mu_b^2}{F_0}.
\end{gathered}
\ee
where in the above $a,b$ run from $1$ to $N-1$.
Had we started from the near-EVH geometry, the AdS$_3$ factor (the $r,t,\psi$ part) of \eqref{NHMPEVH-metric} would have turned into a generic BTZ black hole geometry \cite{NHEVH-1, NHEVH-three-theorems}. The NH geometry \eqref{NHMPEVH-metric} has $SO(2,2)\times U(1)^{N-1}\simeq SL(2,\mathbb{R})\times SL(2,\mathbb{R})\times U(1)^{N-1}$ isometry. This is to be compared with $SL(2,\mathbb{R})\times U(1)^N$ of the non-EVH NHEMP discussed in previous sections.

To discuss separability of the particle dynamics on \eqref{NHMPEVH-metric}, as in the previous sections, we introduce coordinates,
\be
x_a\equiv \frac{a_a \mu_a}{r_0}\,\quad m_a\equiv \frac{a_a^2}{r^2_0}\, \qquad \sum_{a=1}^{N-1}\frac{1}{m_a}=1,
\ee
in which  \eqref{NHMPEVH-metric} takes the form
\be\label{rEVH}
\frac{ds^2}{r^2_0}=F_0ds^2_{AdS_3}+\sum_{a}^{N-1} dx_a^2 +\sum_{a,b}^{N-1} \tilde{\gamma}_{ab}x_ax_bd\varphi_ad\varphi_b,
\ee
with
\be
\begin{gathered}
ds^2_{AdS_3}=
r^2\,\left(-dt^2+{d\psi^2}\right)+\frac{ dr^2}{r^2},
\qquad  F_0=1-\sum_a^{N-1}\frac{x^2_a}{m_a},\\
\tilde{\gamma}_{ab}x_ax_b=\frac{1}{r_0^2}\gamma_{ab},
\qquad \tilde{\gamma}_{ab}=\delta_{ab}+\frac{1}{F_0}\frac{x_a}{\sqrt{m_a}}\frac{x_b}{\sqrt{m_b}}.
\end{gathered}
\ee

The generators of the two $SL(2,\mathbb{R})$ Killing vectors may be written as
\bea
&&H_+={\partial_v}\,,\qquad D_+=v\,{\partial_v}-r\,\partial_r\,\qquad K_+=v^2\,{\partial_v}+\frac{1}{r^2}\,{\partial_u}-2r\, v\,{\partial_r}\,,\nnr
&&H_-={\partial_u}\,,\qquad D_-=u\,{\partial_u}-r\,\partial_r\,\qquad K_-=u^2\,{\partial_u}+\frac{1}{r^2}\,{\partial_v}-2r\, u\,{\partial_r}\,,
\eea
where $v=t+\psi$ and $u=t-\psi$.  The Casimir of $SL(2,\mathbb{R})$'s are
\bea
\mathcal{I_\pm}=H_\pm \,K_\pm-D_\pm^2
\eea
and one can readily check that both Casimirs are equal to $\mathcal{I}=\frac{1}{r^2}\left(\partial_t^2-\partial_{\psi}^2\right)-r^2\,\partial_{r}^2\,.$

The mass-shell equation of the probe particle \eqref{mass-shell-1} then reads
\be
\label{kg}
	\frac{(p_0)^2 -(p_\psi)^2}{r^2}=(rp_r)^2 +\mathcal{I}(p_a,x_a, p_{\varphi_a})
\ee
where
\be
	\{p_a, x_b\}= \delta_{ab},\quad \{p_{\varphi_a},\varphi^b\}=\delta_{ab},\quad \{p_\psi,\psi\}=1, \quad\{p_r, r\}=1,
\ee
and
\be
\begin{gathered}
\label{nullI}
	\mathcal{I}(p_a,x_a, p_{\varphi_a})=
	(1-\sum_{c=1}^{N-1}\frac{x^2_c}{m_c})
	\left[\sum_{a=1}^{N-1}{p^2_a}+\sum_{a=1}^{N-1}\frac{p^2_{\varphi_a}}{x_a^2}+g_0\right],
	\\
	g_0=-\left(\sum_a^{N-1}\frac{p_{\varphi_a}}{\sqrt{m_a}}\right)^2+m_0^2r_0^2,
\end{gathered}
\ee
where  ${\cal I}$ in \eqref{nullI} is the Casimir. Note that while the background has  $SL(2,\mathbb{R})\times SL(2,\mathbb{R})\times U(1)^{N-1}$ isometry the Casimirs of the the two $SL(2,\mathbb{R})$ factors happen to be identically the same and hence we are dealing with a single ${\cal I}$; appearance of an extra $SL(2,\mathbb{R})$ does not add to number of constant of motion compared to the non-EVH case.

Hence, as in the regular case, we have to consider separately three cases

\begin{itemize}
	
	\item {\sl Generic, non-isotropic case, all $m_a$ are non-equal}
	
	To separate the variables in \eqref{nullI}, in the special case when none of the rotational parameter is equal, we introduce the ellipsoidal coordinates
	\bea
	x_a^2=\frac{\prod_{b=1}^{N-1}(m_a-\lambda_b)}{{\prod_{b\ne a}^{N-1}}(m_a-m_b)}.
	\eea
	In this terms the angular Hamiltonian reads
	\be
	\label{eq:vanishing_none_eq}
	\mathcal{I}=\left(\prod_a^{N-1}\frac{\lambda_a}{m_a}\right)
	\left[\sum_{a=1}^{N-1}\frac{4\prod_{b}^{N-1}(m_b-\lambda_a)}{\prod_{b\ne a}^{N-1}(\lambda_b-\lambda_a)}\pi_a^2+\sum_a^{N-1}\frac{p_{\varphi_a}^2}{x_a^2}+g_0\right],
	\ee
	where $\{\pi_a,\lambda_b\}=\delta_{ab}$. One can see that \eqref{eq:vanishing_none_eq} has a very similar form to  \eqref{eq:conf_ham_odd}, and using the identities  \eqref{26} and \eqref{eq:relation_1}, it can be rewritten as follows (after fixing the Hamiltonian $\mathcal{I}=\mathcal{E}$)
	\be	
	\label{eq:vanishing_R}
	\sum_{a=1}^{N-1}\frac{R_a- \mathcal{\tilde{E}}}{\lambda_a\prod_{b=1,a\ne b}^{N-1}(\lambda_b-\lambda_a)}=0,
	\ee
	where
	\be	
	\begin{gathered}
	R_a=4\lambda_a\pi_a^2\prod_{b}^{N-1}(m_b-\lambda_a)+(-1)^{N-1}\sum_b^{N-1}\frac{\lambda_a g_b^2}{\lambda_a-m_b}-g_0(-\lambda_a)^{N-1},\\
	g_a^2=p_{\varphi_a}^2\prod_{b=1}^{N-1}(m_a-m_b),\qquad
	\mathcal{\tilde{E}}=\mathcal{E}\prod_a^{N-1}m_a.
	\end{gathered}
	\ee
	Separation of variables and the constants of motion is similar to the \sect{\ref{Particle-dynamics-Sec}}, where \eqref{eq:vanishing_R} corresponds to \eqref{HJO}.
	
	\item {\sl Isotropic case, all $m_a$ are equal}
	
	In this case ($m_a=N-1$), we separate the variables in \eqref{nullI} by introducing spherical coordinates $\{u,\ ,y_\alpha,\ \theta_{N-2}\}$
	\be
	x_{N-1}=u \cos \theta_{N-2}, \qquad x_{N-1-\alpha}=u\ y_\alpha \sin \theta_{N-2},\qquad
	\sum_{\alpha=1}^{N-2}y_\alpha ^2=1
	\ee
	where $\alpha=1\ldots N-2$. In these coordinates \eqref{nullI} will take the following form
	\be
	\mathcal{I}=\left(1-\frac{1}{N-1}u^2\right)\left[p_u^2+\frac{F_{N-2}}{u^2}+g_0\right]
	\ee
	with $F_a$ defined in \eqref{eq:iso_integrals_odd}, where the separation of variables and the derivation of integrals of motion was carried out according to \ref{fully-isotropic-sec}.
	

	\item {\sl partially isotropic case}
	
	The last case is the most general one which involves  sets of equal  and a set of non-equal rotational parameters. With the discussions of the two previous cases (fully isotropic and fully non-isotropic) in view and  recalling the analysis of partially isotropic NHEMP case of previous section, it is  straightforward to separate the variables in partially isotropic NHEVHMP. Following the steps in \ssect{\ref{sec:gc}}, one should first introduce different spherical coordinates for each set of equal rotational parameters and ellipsoidal coordinates for the joint set of non-equal rotational parameters and the radial parts of spherical coordinates. This will result into a spherical mechanics similar to \eqref{eq:red_mech_inter} where the Hamiltonians of spherical subsystems will be included as parameters.
	
\end{itemize}

\subsection{Discussion}
Continuing the analysis of \cite{non-equal-general, Demirchian:2017uvo}, we studied  separability of geodesic motion on the near horizon geometries of Myers-Perry black hole in $d$, even or odd, dimensions and established the integrability by explicit construction of $d$ constants of motion.  In the general case $[\frac{d-1}{2}]+1$ of these constants of motion are related to the Killing vectors of the background (note that the background in general has $[\frac{d-1}{2}]+3$ Killing vectors, but three of them form an $sl(2,\mathbb{R})$ algebra and hence there is only one independent  conserved charge from this sector). Our analysis reconfirms the earlier observations that although near-horizon limit in the extremal black holes enhances the number of Killing vectors by two \cite{NHEG-general}, the number of independent conserved charges from the Killing vectors does not change.
Our system, in the general case,  has $[\frac{d}{2}]$ constants of motion associated with second rank Killing tensors the system possesses. We also constructed  the explicit relation between these Killing tensors  and the conserved charges and one may check that our Killing tensors and those in  \cite{Hidden-symmetry-NHEK} match. We note that the Killing tensors of \cite{Hidden-symmetry-NHEK} were obtained using the near horizon limit  on the Killing tensors of Myers-Perry black hole in a coordinate system which makes the geodesics of black hole separable itself. Whereas, we directly worked with ellipsoidal coordinates for the NHEMP,  introduced in \cite{non-equal-general}. Comparing the two systems before and after the NH limit,
it was argued in  \cite{Hidden-symmetry-NHEK} that a combination of Killing tensors is reducible to the Killing vectors, however, we obtain other second rank Killing tensors, through which the system remains integrable.
Moreover, by explicitly showing the separability, one concludes that  there is no inconsistency with the theorems in \cite{Benenti and Francaviglia}. There is an extra conserved charge related to the Casimir of $SL(2,\mathbb{R})$ symmetry group which intrinsically exists in the NHEG's. We have shown that the charge of the Casimir is independent of the other conserved charges. In this sense, one of the ``hidden symmetries,'' symmetries which are associated with equations of motion and are not isometries of the background,  becomes explicit in the NH limit \cite{Hidden-symmetry-NHEK}.

Following the discussions in \cite{GNS-11,GNS-12}, we showed that for special cases where some of the rotations parameters of the background are equal, the  geodesic problem on NHEMP is superintegrable. We established superintegrability by establishing existence of  other constants of motion.  Our methods here, combined with those in \cite{GNS-11,GNS-12}, allows one to read the extra second rank Killing tensors obtained in these cases. The rough picture is as follows: We started with a system with $2N+1+\sigma$ variables with $N$ isometries. Fixing the momenta associated with the isometries, we obtained and focused the $N-1+\sigma$ dimensional ``angular mechanics'' part.  In this sector, whenever $N$ number of rotation parameters $m_i$ of the background metric are equal the $U(1)^N$ isometry is enhanced to $U(N)$ and this latter brings about other second rank Killing tensors. All in all, the fully isotropic case in odd dimensions with $U(\frac{d-1}{2})$ isometry, the $d-2$ dimensional spherical mechanics part  is maximally superintegrable, it has $N+(N-2)=2N-2$ extra constants of motion. The fully isotropic case in even dimensions, however, is not maximally superintegrable; it has still $2N-1$  extra Killing tensors (one less than the $N$ constants of motion to make the system fully superintegrable).  We discussed the ``special cases'' in two different ways. First, we reanalyzed the system from the scratch (in \sect{\ref{fully-isotropic-sec}}) and also took the equal rotation parameter limit of the generic case (in \ssect{\ref{sec:contrraction}}). As expected, these two cases matched. Our preliminary analysis, which we did not show here, indicate that the above statements is also true for the NH limit of extremal MP black holes in (A)dS backgrounds.

We also discussed the EVH case, which happens for odd dimensional extremal MP when one of the  rotation parameters $a_i$ vanishes. In the general NHEVHMP case, where the background isometry is $SO(2,2)\times U(1)^{\frac{d-3}{2}}$ the number of independent charges associated with Killing vectors is $\frac{d+1}{2}$. Despite enhancement of the isometry group compared to the generic NHEMP case, we found that this symmetry enhancement does not add independent constants of motion, the system in general does not poses extra constants of motion and remains just integrable.

Here we explored second rank Killing tensors. One may suspect that the system has independent higher rank Killing tensors too, although it is unlikely. But if it does, the system for the generic rotation parameters becomes superintegrable. It is interesting to explore this question.
Finally, as already pointed out in the introduction, one can consider other probes including scalar, Dirac field or gauge or tensor perturbations on the NHEMP backgrounds and study their integrability. To this end, the study of Killing Yano tensor and principal tensor \cite{Frolov:2007nt,Kubiznak:2006kt} should be completed.


%% file: Klein_Gordonization.tex
\section[Superintegrable quantum systems and resonant spacetimes]
				{Superintegrable quantum systems 
			\\ and resonant spacetimes}
\label{sec:klein_gordon}
Geometrization of dynamics is a recurrent theme in theoretical physics. While it has underlied such fundamental developments as the creation of General Relativity and search for unified theories of interactions, it also has a more modest (but often fruitful) aspect of reformulating conventional, well-established theories in more geometrical terms, in hope of elucidating their structure. One particular approach of the latter type is the Jacobi metric (for a contemporary treatment, see \cite{jacobi1,jacobi2,jacobi3}). This energy-dependent metric simply encodes as its geodesics the classical orbits of a nonrelativistic mechanical particle on a manifold moving in a potential.

The geometrization program we propose here can be seen as a quantum counterpart of the Jacobi metric. To a nonrelativistic quantum particle on a manifold moving in a potential, we shall associate a relativistic Klein-Gordon equation in a static spacetime of one dimension higher. Since the Klein-Gordon equation can be seen as a sort of quantization of geodesics (and reduces to the geodesic equation in the eikonal approximation), this provides a quantized version of the correspondence between particle motion on a manifold in the presence of a potential and purely geometric geodesic motion in the corresponding spacetime. Executing our geometrization algorithm in general reduces to a nonlinear elliptic equation closely reminiscent of the one emerging in relation to the Yamabe problem and its generalizations known as prescribed scalar curvature problems \cite{yamabe1,yamabe2,prescr}, and thus connects to extensive literature and interesting questions in differential geometry. (The Yamabe problem refers to constructing a conformal transformation of the given metric on a manifold that makes the Ricci scalar of the conformally transformed metric constant.)

While the correspondence we build may in principle operate on any system, we are primarily motivated by its application to a very special class of quantum systems whose energy is a quadratic function of the energy level number. Such systems are exemplified by the one-dimensional P\"oschl-Teller potential, and in higher dimensions they are typically superintegrable. In fact, our construction has been developed precisely as a generalization of the correspondence \cite{EK,EN,EN2} between the Higgs oscillator \cite{Higgs,Leemon}, a particularly simple superintegrable system with a quadratic spectrum, and Klein-Gordon equations on the Anti-de Sitter (AdS) spacetime, the maximally symmetric spacetime of constant negative curvature. This correspondence has emerged in the context of studying selection rules \cite{CEV1,CEV2,Yang} in the nonlinear perturbation theory targeting the AdS stability problem \cite{BR,review}. The correspondence has been useful for both elucidating the structure of AdS perturbation theory \cite{EN} and for resolving the old problem of constructing explicit hidden symmetry generators for the Higgs oscillator \cite{EN2}.

The reason for our emphasis on systems with quadratic spectra is that, in application to such systems, our geometrization program generates Klein-Gordon equations whose frequency spectra are linear in the frequency level number, and hence the spectrum is highly resonant (the difference of any two frequencies is integer in appropriate units). It is well-known that in the context of weakly nonlinear dynamics, highly resonant spectra have a dramatic impact, as they allow arbitrarily small nonlinear perturbations to produce arbitrarily large effects over long times. This feature has been crucial in the extensive investigations of the AdS stability problem in the literature (for a brief review and references, see \cite{review}). The main practical target of our geometrization program thus appears twofold:
\begin{itemize}
	\item to provide geometric counterparts for quantum systems with quadratic spectra (the resulting Klein-Gordon equation is set up on a highly special spacetime with a resonant spectrum of frequencies and the geometric properties of this spacetime are likely to yield insights into the algebraic properties of the original quantum system, including its high degree of degeneracy and hidden symmetries),
	\item to generate, starting from known quantum systems with quadratic spectra, highly resonant spacetimes (weakly nonlinear dynamics in such spacetimes is likely to be very sophisticated, sharing the features of the extensively explored weakly nonlinear dynamics of AdS).
\end{itemize}

In \sect{\ref{subsec:general_formulation}} and \sect{\ref{subsec:massless_case}}, we formulate our general geometrization procedure and describe how it simplifies for the case of zero mass in the Klein-Gordon equation one is aiming to construct. In \sect{\ref{subsec:higgs}}, we describe how the previously known correspondence \cite{EK,EN,EN2} between the Higgs oscillator and AdS fits in our general framework. In \sect{\ref{subsec:rosochatius}}, we analyze the superintegrable Rosochatius system, which generalizes the Higgs oscillator, and generate a large family of spacetimes perfectly resonant with respect to the massless wave equation. We conclude with a review of the current state of our formalism and open problems.

The results of this chapter were obtained in cooperation with Oleg Evnin and are based on \cite{Evnin:2017vpc}.

\subsection{General formulation of the procedure}
\label{subsec:general_formulation}
Consider a quantum system with the Hamiltonian
\beq
H=-\Delta_\gamma + V(x),
\label{Hamlt}
\eeq
where $\Delta_\gamma\equiv \gamma^{-1/2}\del_i(\gamma^{1/2}\gamma^{ij}\del_j)$ is the Laplacian on a $d$-dimensional manifold parametrized with $x^i$ and endowed with the metric $\gamma_{ij}$. We shall be particularly interested in systems whose energy spectrum consists of (in general, degenerate) energy levels labelled by the level number $N=0$, 1, ..., and the energy is a quadratic function of the level number:
\beq
E_N= A(N+B)^2-C.
\label{quadrenergy}
\eeq
Such spectra are indeed observed in a number of interesting systems, typically involving superintegrability, for example:
\begin{itemize}
	\item The Higgs oscillator \cite{Higgs,Leemon}, which is a particle on a $d$-sphere moving in a potential varying as the inverse cosine-squared of the polar angle.
	\item The superintegrable version \cite{rossup1,rossup2} of the Rosochatius system on a $d$-sphere \cite{rosochatius,encycl}, which is the most direct generalization of the Higgs oscillator.
	\item The quantum angular Calogero-Moser model \cite{FLP}.
	\item The (spherical) Calogero-Higgs system \cite{HLN,CHLN}.
\end{itemize}
We additionally mention the following two completely elementary systems which give a particularly simple realization of the quadratic spectrum (\ref{quadrenergy}):
\begin{itemize}
	\item A particle in one dimension in an infinite rectangular potential well.
	\item The trigonometric P\"oschl-Teller system \cite{PT}. 
\end{itemize}

We would like to associate to any system of the form (\ref{Hamlt}) a Klein-Gordon equation in a certain static $(d+1)$-dimensional space-time. We introduce a scalar field $\tilde\phi(t,x)$ satisfying
\beq\label{secondorder}
-\del_t^2\tilde\phi=(-\Delta_\gamma+V(x)+C)\tilde\phi.
\eeq
In the above expression, $C$ can in principle be an arbitrary constant, but our main focus will be on systems with energy spectrum of the form (\ref{quadrenergy}) and $C$ read off from (\ref{quadrenergy}). One can equivalently write (\ref{secondorder}) as
\beq\label{KGwannabe}
\Box_{\tilde g} \tilde\phi -(C+V(x))\tilde\phi=0.
\eeq
Where $\Box_{\tilde g}$ is the D'Alembertian of the metric
\beq\label{tildeg}
\tilde g_{\mu\nu}dx^\mu dx^\nu=-dt^2+\gamma_{ij}dx^i dx^j,
\eeq
with $x^\mu=(t,x^i)$. By construction, if one implements separation of variables in (\ref{KGwannabe}) in the form
\beq
\tilde\phi=e^{i w t}\Psi(x),
\eeq
one recovers the original Schr\"odinger equation as $H\Psi=(w^2-C)\Psi$. This guarantees that the mode functions of (\ref{KGwannabe}) are directly related to the energy eigenstates of the original quantum-mechanical problem.
Note that, if one focuses on systems with energy spectra of the form (\ref{quadrenergy}), by construction, separation of variables in (\ref{secondorder}) will lead to eigenmodes with linearly spaced frequencies:
\beq\label{omegaN}
w_N=\sqrt{A}(N+B).
\eeq
In this case, after conversion to the Klein-Gordon form, which we shall undertake below, the resulting spacetime will possess a  resonant spectrum of frequencies.

Equation (\ref{secondorder}) is not of a Klein-Gordon form, but we can try to put in this form by applying a conformal rescaling to $\tilde g$ and $\tilde\phi$:
\beq
\label{eq:conf_resc}
\tilde g_{\mu\nu}=\Omega^2 g_{\mu\nu},\qquad \tilde\phi=\Omega^{\frac{1-d}2}\phi.
\eeq
One thus gets (relevant conformal transformation formulas can be retrieved, e.g., from \cite{BD})
\beq
\Box_g\phi - \left[(C+V(x))\Omega^2+\frac{d-1}2\frac{\Box_g\Omega}{\Omega}+\frac{(d-1)(d-3)}4\frac{g^{\mu\nu}\del_\mu\Omega\del_\nu\Omega}{\Omega^2}\right]\phi=0.
\eeq
If the expression in the square brackets can be made constant by a suitable choice of $\Omega$, we get a Klein-Gordon equation in a spacetime with the metric $g_{\mu\nu}$. We thus need to solve the equation
\beq\label{conformal}
(C+V(x))\Omega^2+\frac{d-1}2\frac{\Box_g\Omega}{\Omega}+\frac{(d-1)(d-3)}4\frac{g^{\mu\nu}\del_\mu\Omega\del_\nu\Omega}{\Omega^2}=m^2.
\eeq
It is wiser to rewrite this equation through the metric $\tilde g$, which is already known and given by (\ref{tildeg}):
\beq
\frac{d-1}2\Omega{\Box_{\tilde g}\Omega}-\frac{d^2-1}4\tilde g^{\mu\nu}\del_\mu\Omega\del_\nu\Omega+(C+V(x))\Omega^2=m^2.
\eeq
Since neither $V(x)$ nor $\tilde g_{\mu\nu}$ depend on $t$, one can assume that $\Omega$ is a function of $x^i$ as well. Hence,
\beq\label{confgamma}
\frac{d-1}2\Omega{\Delta_{\gamma}\Omega}-\frac{d^2-1}4\gamma^{ij}\del_i\Omega\del_j\Omega+(C+V(x))\Omega^2=m^2.
\eeq

Note that (\ref{conformal}) is closely reminiscent of the equation emerging from the following purely geometrical problem: Consider a metric $g_{\mu\nu}$ whose Ricci scalar is $R(x)$. Is it possible to find $\Omega$ such that the Ricci scalar corresponding to $\tilde g_{\mu\nu}=\Omega^2 g_{\mu\nu}$ has a given form $\tilde R(x)$? Indeed, from the standard formulae for the change of the Ricci scalar under conformal transformations, see, e.g., (3.4) of \cite{BD}, one gets
\beq\label{rrtilde}
\Omega^2 \tilde R(x)= R(x)+2d\frac{\Box_g\Omega}{\Omega}+d(d-3)\frac{g^{\mu\nu}\del_\mu\Omega\del_\nu\Omega}{\Omega^2}.
\eeq
Algebraically, this has the same structure as (\ref{conformal}).

Equations of the form (\ref{rrtilde}) for simple specific choices of $g$ and $\tilde R$ have been studied in mathematical literature as various realizations of the `prescribed scalar curvature' problem \cite{prescr}. Substitution
\beq
\Omega=\omega^{-\frac2{d-1}},
\eeq
reduces (\ref{confgamma}) to the following compact form
\beq
-\Delta_\gamma \omega+(C+V(x))\omega=m^2\omega^\frac{d+3}{d-1},
\label{yma}
\eeq
closely reminiscent to the equation arising in relation to the Yamabe problem \cite{yamabe1,yamabe2,prescr}. (Note that the specific power of $\omega$ appearing on the righ-hand side of this equation is different from the standard Yamabe problem. This is because we are performing a conformal transformation in a spacetime of one dimension higher, rather than in the original space.) Once (\ref{yma}) has been solved, the spacetime providing geometrization of the original problem (\ref{Hamlt}) can be written explicitly as
\beq
g_{\mu\nu}dx^\mu dx^\nu=\omega^{\frac{4}{d-1}}\left(-dt^2+\gamma_{ij}dx^i dx^j\right).
\label{gsol}
\eeq

Equation (\ref{confgamma}) dramatically simplifies in one spatial dimension ($d=1$), where all the derivative terms drop out, leaving $\Omega\sqrt{C+V(x)}=m$. Thus, for the particle in an infinite rectangular potential well, Klein-Gordonization gives a massless wave equation on a slice of Minkowski space between two mirrors, while for the P\"oschl-Teller system, one immediately obtains a two-dimensional spacetime metric reminiscent of Anti-de Sitter spacetime AdS$_2$. This latter result displays some parallels to the considerations of \cite{CJP} (focusing in the hyperbolic P\"oschl-Teller system).

As we already briefly remarked, the above geometrization procedure can be applied to any Hamiltonian of the form (\ref{Hamlt}) and any $C$, irrespectively of the form of the spectrum. However, it is precisely for the spectrum and $C$ given by (\ref{quadrenergy}) that the resulting spacetime possesses the remarkable property of being highly resonant (and one may expect that its geometric properties will give a more transparent underlying pictures of the algebraic structurs of the original quantum-mechanical problem, as happens for the Higgs oscillator). We shall therefore focus on the application of our geometrization procedure to such systems with quadratic energy spectra.

\subsection{The massless case}\label{massless}
\label{subsec:massless_case}
Equation (\ref{yma}) is a nonlinear elliptic equation and in general difficult to solve. Extensive existence result are established for an algebraically similar equation arising in relation to the Yamabe problem, hence one may hope that some level of understanding of solutions to (\ref{yma}) in full generality may also be attained in the future. We shall not pursue such systematic analysis here, however.

Driven by practical goals of constructing resonant spacetimes and geometrizing concrete superintegrable systems, we would like to point out  that (\ref{yma}) becomes linear and dramatically simplifies if one assumes $m^2=0$. Hence, converting a given quantum mechanical problem to a massless wave equation is considerably simpler than for general values of the mass.

We note that, if $m^2=0$, equation (\ref{yma}) looks identical to the Schr\"odinger equation corresponding to the Hamiltonian (\ref{Hamlt}), with energy eigenvalue $-C$:
\beq
-\Delta_\gamma \omega+V(x)\omega=-C\omega,
\label{yma0}
\eeq
(Normalizable eigenstates of this energy do not generically exist, but $\omega$ does not have to satisfy the same normalizability conditions as standard wave functions, hence this should not be a problem.) Since quadratic spectra (\ref{quadrenergy}) are seen to arise from highly structured, typically superintegrable, systems, one may naturally expect that (\ref{yma0}) is amenable to analytic treatment.

There is one further assumption one might make that immediately yields solutions of (\ref{yma0}) from known solutions of the original quantum-mechanical problem (\ref{Hamlt}). Namely, imagine one has an $K$-parameter family of Hamiltonians (\ref{Hamlt}) with quadratic spectra (\ref{quadrenergy}). In this case, $A$, $B$, and $C$ are functions of the $K$ parameters defining our family of Hamiltonians. One may impose
\beq
B=0,
\eeq
which generically yields an $(K-1)$-parameter subfamily of quantum systems with quadratic spectra. Within this subfamily, the ground state $\Psi_0$ has the energy $-C$, i.e., $H\Psi_0=-C\Psi_0$. Hence, $\omega$ satisfying (\ref{yma0}) can be chosen as the vacuum state of $H$:
\beq
\omega=\Psi_0.
\label{omegaPsi}
\eeq
We shall make use of this construction below, as it allows for a straightforward application of our methodology to known exactly solvable systems. (In some cases, it is geometrically advantageous to use the non-normalizable counterpart of $\Psi_0$ with the same energy eigenvalue to define $\omega$. Such non-normalizable states should also be easy to construct for exactly solvable systems with quadratic spectra. We shall see an explicit realization of this scenario in our subsequent treatment of the superintegrable Rosochatius problem.)

As a variation of the above special case, one could force $B$ of (\ref{quadrenergy}) to be equal to a negative integer and $\omega$ to be equal to an excited state wavefunction. This, however, introduces singularities in the conformally rescaled spacetime (\ref{gsol}) at the location of zeros of the excited state wavefunctions. While one could still try to pursue this scienario by imposing appropriate constraints on the wave equation solution at the singular locus, we shall concentrate below on the most straightforward formulation (\ref{omegaPsi}) utilizing the ground state wavefunction, where the conformal factor is non-vanishing and no such subtleties arise.

\subsection{KG background dual to Higgs oscillator}
\label{subsec:higgs}
Before proceeding with novel derivations we would like to demonstrate how the case of the Higgs oscillator, which has motivated our general construction, fits into our present framework. We are essentially just reviewing the derivations in \cite{EK,EN,EN2}.

The Higgs oscillator is a particle on a sphere moving in a specific centrally symmetric potential (which we shall specify below). It is remarkable for being one of only three centrally symmetric maximally superintegrable systems on a sphere (together with free motion and the spherical Coulomb potential). A practical manifestation of superintegrability is that all of its classical trajectories are closed. The quantum version of this system has attracted considerable attention after it was reintroduced in a different guise and solved in \cite{ES}. The observed high degeneracy of energy levels of this system prompted investigation of its hidden symmetries in \cite{Higgs,Leemon}, which resulted in identification of the hidden $SU(d)$ group of symmetries for a system on a $d$-sphere, and spawned extensive literature on algebras of conserved quantities of the Higgs oscillator.  The energy spectum of the Higgs oscillator is of the form (\ref{quadrenergy}).

We shall now define, with some geometric preliminaries, the Higgs oscillator Hamiltonian. Consider a unit $d$-sphere embedded in a (d+1)-dimensional flat space as
\beq
x_0^2+x_1^2+\cdots+x_{d}^2=1
\eeq
and parametrized by the angles $\te_1$, ..., $\te_{d}$ as
\begin{align}\label{xsphere}
&x_{d}=\cos\te_{d},\\
& x_{d-1}=\sin\te_{d}\cos\te_{d-1},\\ \nonumber
&x_1=\sin\te_{d}\ldots\sin\te_2\cos\te_1,\qquad \\\nonumber
&x_0=\sin\te_{d}\ldots\sin\te_2\sin\te_1.\nonumber
\end{align}
The sphere is endowed with the standard round metric defined recursively in $d$
\begin{align}
&ds^2_{S^d}=d\te_{d}^2+\sin^2\te_{d}ds^2_{S^{d-1}},\\
 &ds^2_{S^1}=d\te_{1}^2. \nonumber
\end{align}
Similarly, the corresponding Laplacian is defined recursively
\begin{align}\label{Deltasphere}
&\Delta_{S^d}=\frac1{\sin^{d-1}\te_{d}}\del_{\te_{d}}\left(\sin^{d-1}\te_{d}\,\del_{\te_{d}}\right)+\frac1{\sin^2\te_{d}}\Delta_{S^{d-1}},\\
 &\Delta_{S^1}=\del^2_{\te_1}.\nonumber
\end{align}

The Higgs oscillator is a particle on a $d$-sphere moving in a potential varying as the inverse cosine-squared of the polar angle:
\beq
H=-\Delta_{S^d}+\frac{\alpha(\alpha-1)}{\cos^2\te_d}.
\label{higgsH}
\eeq
The energy spectrum is given by
\beq\label{Higgsenrg}
E_N=\left(N+\alpha+\frac{d-1}2\right)^2-\frac{(d-1)^2}4,
\eeq
where $N$ is the energy level number. This expression is manifestly of the form (\ref{quadrenergy}).

To implement our geometrization program for the Higgs oscillator, one can work directly with (\ref{confgamma}), which takes the form
\beq
\frac{d-1}2\frac{\Omega}{\sin^{d-1}\te_d}\del_{\te_d}(\sin^{d-1}\te_d\,\del_{\te_d} \Omega)-\frac{d^2-1}4(\del_{\te_d}\Omega)^2+\left(C+\frac{\alpha(\alpha-1)}{\cos^2\te_d}\right)\Omega^2=m^2.
\eeq
Substituting $\Omega=\cos\te_d$ produces only two constraints on the parameters to ensure that the equation is satisfied:
\beq
C=\frac{(d-1)^2}4,\qquad m^2=\alpha(\alpha-1)+\frac{d^2-1}4.
\eeq
The value of $C$ above agrees with the one in (\ref{Higgsenrg}), while the relation between the Klein-Gordon mass and the Higgs potential strength is the same as found in \cite{EK}. The output of our construction is thus a family of Klein-Gordon equations on the spacetime
\beq\label{AdSHiggs}
ds^2=\frac{-dt^2+ds^2_{S^d}}{\cos^2\te_d},
\eeq
which is precisely the (global) Anti-de Sitter spacetime AdS$_{d+1}$. We note that rational values of $\alpha$ in (\ref{higgsH}) correspond to Klein-Gordon masses in AdS for which the frequency spectrum (\ref{omegaN}) is perfectly resonant (all frequencies are integer in appropriate units) rather than merely strongly resonant (differences of any two frequencies are integer in appropriate units).

A remarkable property of the Higgs oscillator is that the metric (\ref{AdSHiggs}) does not depend on the Higgs potential strength (which only affects the value of the Klein-Gordon mass). This feature is not replicated for more complicated potentials. Conversely, this implies that the AdS spacetime possesses a resonant spectrum of frequencies for fields of all masses (this statement can in fact be extended to fields of higher spins), rather than for fields of one specific mass. It is tempting to conjecture that AdS (being a maximally symmetric spacetime) is the only spacetime with this property, though we do not know a proof. Relations between Klein-Gordon equations of different masses have recently surfaced in the literature on ``mass ladder operators'' \cite{mass1,mass2,mass3,mass4}.

\subsection{KG background dual to  Rosochatius system}
\label{subsec:rosochatius}
The superintegrable Rosochatius system is the most direct generalization of the Higgs oscillator on a $d$-sphere preserving its superintegrability. General Rosochatius systems \cite{rosochatius} were among the first Liouville-integrable systems discovered. A restriction on the potential makes these systems maximally superintegrable. The Higgs oscillator can be recovered by a further restriction of the potential as a particularly simple special case. Such systems are thus an ideal testing ground for applying our machinery, which has already been shown to work for the Higgs oscillator.
\subsubsection{The superintegrable Rosochatius system}

The superintegrable Rosochatius systems we shall deal with here are defined by the following family of Hamiltonians:
\beq
H^{(R)}_d=-\Delta_{S^d}+\sum_{k=0}^{d}\frac{\alpha_k(\alpha_k-1)}{x_k^2}.
\label{rosH}
\eeq
The explicit form of the Laplacian and coordinates on the unit $d$-sphere can be read off from (\ref{xsphere}-\ref{Deltasphere}). The standard more general definition of the Rosochatius system \cite{rosochatius,encycl} additionally includes a harmonic potential with respect to the $x_k$ variables, $\sum_k \gamma_kx_k^2$, which gives an integrable system. If this harmonic potential is omitted, as we did above, the system becomes maximally superintegrable, as mentioned, for instance, in \cite{rossup1,rossup2}.

In order to find the spectrum of the above Hamiltonian, we shall have to apply recursively the solution of the famed one-dimensional P\"oschl-Teller problem \cite{PT}. While this material is completely standard and occasionally covered in textbooks, we find the summary given in \cite{IH} concise and convenient. The energy eigenstates of the P\"oschl-Teller Hamiltonian
\beq
H_{PT}=-\del_x^2+\frac{\mu(\mu-1)}{\cos^2x}+\frac{\nu(\nu-1)}{\sin^2x}
\label{HPT}
\eeq
are given by
\beq
\eps_n=(\mu+\nu+2n)^2,\qquad n=0,1,2,\cdots
\label{PTenergy}
\eeq
We shall not need the explicit form of the eigenfunctions satisfying $H_{PT}\Psi_n=\eps_n\Psi_n$ (though it is known).

Because of the recursion relations on $d$-spheres outlined above, the Rosochatius Hamiltonian (\ref{rosH}) can likewise be defined recursively:
\begin{align}
&H^{(R)}_d=-\frac1{\sin^{d-1}\te_{d}}\del_{\te_{d}}\left(\sin^{d-1}\te_{d}\,\del_{\te_{d}}\right)+\frac{\alpha_d(\alpha_d-1)}{\cos^2\te_d}+\frac{1}{\sin^2\te_d}H^{(R)}_{(d-1)},\\
&H^{(R)}_1=-\del^2_{\te_1}+\frac{\alpha_1(\alpha_1-1)}{\cos^2\te_1}+\frac{\alpha_0(\alpha_0-1)}{\sin^2\te_1}.
\end{align}
The variables separate, and if one substitutes the wave function in the form
\beq
\Psi(\te_1,\cdots,\te_d)=\prod_{p=1}^{d}\frac{\chi_p(\te_p)}{\sin^{(p-1)/2}\te_p},
\label{sepvar}
\eeq
one obtains a recursive family of one-dimensional eigenvalue problems, all of which are of the P\"oschl-Teller form:
\begin{align}\label{varsep}
&\left[-\del_{\te_d}^2+\frac{\alpha_d(\alpha_d-1)}{\cos^2\te_d}+\left(\frac{(d-2)^2-1}4+E_{d-1}\right)\frac1{\sin^2\te_d}-\frac{(d-1)^2}4\right]\chi_d=E_d\chi_d,\\
&\left[-\del^2_{\te_1}+\frac{\alpha_1(\alpha_1-1)}{\cos^2\te_1}+\frac{\alpha_0(\alpha_0-1)}{\sin^2\te_d}\right]\chi_1=E_1\chi_1,\nonumber
\end{align}
where $E_d$ are eigenvalues of $H^{(R)}_d$. Each subsequent equation introduces one new quantum number which we shall denote $n_d$.

The recursive solution of (\ref{varsep}) proceeds as follows. First, the solution at $d=1$ is given by (\ref{PTenergy}) as
\beq
E_1(n_1)=(\alpha_0+\alpha_1+2n_1)^2.
\eeq
At $d=2$, one gets
\beq
\left[-\del_{\te_2}^2+\frac{\alpha_2(\alpha_2-1)}{\cos^2\te_2}+\frac{(\alpha_0+\alpha_1+2n_1+\frac12)(\alpha_0+\alpha_1+2n_1-\frac12)}{\sin^2\te_2}-\frac{1}4\right]\chi_2=E_2\chi_2.
\eeq
Hence,
\beq
E_2(n_1,n_2)=\left(\alpha_0+\alpha_1+\alpha_2+2n_1+2n_2+\frac12\right)^2-\frac14.
\eeq
The general pattern can now be guessed as
\beq
E_d(n_1,\cdots,n_d)=\left(\alpha_0+\cdots+\alpha_d+2n_1+\cdots+2n_d+\frac{d-1}2\right)^2-\frac{(d-1)^2}4.
\label{rosenergy}
\eeq
It is straightforward to prove inductively that this expression persists under the recursion given by (\ref{varsep}). Note that (\ref{rosenergy}) is manifestly of the form (\ref{quadrenergy}). A classical version of the same construction, recursively expressing the superintegrable Rosochatius Hamiltonian through the action-angle variables has been given in \cite{rossup2}.

\subsubsection{The dual KG background}

To demonstrate how the geometrization procedure we have proposed above operates, we shall now apply it to the superintegrable Rosochatius system. For the purposes of demonstration, we shall use the simplest formulation outlined in \sect{\ref{massless}}, which allows one to utilize known explicit solutions for ground state wavefunctions to construct the relevant massless Klein-Gordon (wave) equation.

The only technical input we shall need is the form of the ground state wavefunction of the P\"oschl-Teller Hamiltonian (\ref{HPT}) given by
\beq
\psi_0=\cos^\mu x\sin^\nu x.
\label{psi0}
\eeq
(This form satisfies the standard boundary conditions for physical wavefunctions only for $\mu\ge0$ and $\nu\ge0$. If not, $\mu$ must be replaced by $1-\mu$, and correspondingly for $\nu$. This is, however, completely irrelevant for our application of $\psi_0$ to construct geometrical conformal factors, and the above form, without any modifications, is perfectly suitable for our purposes.) From (\ref{psi0}) and the recursive construction (\ref{sepvar}-\ref{rosenergy}), one gets for the ground state wavefunction of the superintegrable Rosochatius Hamiltonian (\ref{rosH})
\beq
\Psi_0(\te_1,\cdots,\te_d)=\prod_{p=1}^d\left[\left(\cos\te_p\right)^{\alpha_p}\left(\sin\te_p\right)^{\alpha_0+\alpha_1+\cdots+\alpha_{p-1}}\right].
\eeq
On the other hand, $B$ defined by (\ref{quadrenergy}) can be read off (\ref{rosenergy}) as
\beq
B=\alpha_0+\alpha_1+\cdots+\alpha_d+\frac{d-1}2.
\eeq
We can hence directly apply the algorithm of \sect{\ref{massless}} by introducing
\beq
\omega=\prod_{p=1}^d\left[\left(\cos\te_p\right)^{\alpha_p}\left(\sin\te_p\right)^{\alpha_0+\alpha_1+\cdots+\alpha_{p-1}}\right].
\eeq
under the assumption that
\beq
\alpha_0+\alpha_1+\cdots+\alpha_d+\frac{d-1}2=0.
\eeq
This yields a $d$-parameter family of spacetimes given by (\ref{gsol}) whose massless wave equations possess perfectly resonant spectra and geometrize the superintegrable Rosochatius problem:
\beq
ds^2=\omega^{\frac{4}{d-1}}\left(-dt^2+ds^2_{S^d}\right).
\label{rosmetr}
\eeq
(Note that setting $\alpha_d=-(d-1)/2$ and the rest of $\alpha_p$ to 0 returns the case of Higgs oscillator with the coupling strength corresponding to zero mass in the Klein-Gordon equation, while (\ref{rosmetr}) becomes the AdS metric.)

For a final statement of our result, it is convinient to reparametrize $\alpha_p$ as
\begin{align}
\begin{split}
\alpha_p&=-\frac{d-1}2\beta_p\quad\mbox{for}\quad p\ge 1,\\
 \alpha_0&=-\frac{d-1}2\left(1-\beta_1-\cdots-\beta_d\right).
\end{split}
\end{align}
In terms of $\beta_p$, (\ref{rosmetr}) becomes
\beq
ds^2=
\frac{-dt^2+ds^2_{S^d}}{\dsty\prod_{p=1}^d\left[\left(\cos\te_p\right)^{2\beta_p}\left(\sin\te_p\right)^{2(1-\beta_p-\cdots-\beta_{d})}\right]}.
\label{rosmetr_final}
\eeq
This evidently agrees with (\ref{AdSHiggs}) when $\beta_d=1$ and the rest of $\beta_p$ are zero.

\subsection{Discussion}

We have presented a procedure (``Klein-Gordonization'') associating to quantum systems of the form (\ref{Hamlt}) a Klein-Gordon equation on a static spacetime given by (\ref{gsol}). For systems with the quadratic energy spectrum (\ref{quadrenergy}), our procedure results in spacetimes with a resonant spectrum of evenly spaced frequencies (\ref{omegaN}). This correspondence generalizes the previously known relation between the Higgs oscillator (\ref{higgsH}) and (global) Anti-de Sitter spacetime (\ref{AdSHiggs}).

Implementing our procedure in practice requires solving a nonlinear elliptic equation, which can be written as (\ref{confgamma}) or (\ref{yma}). The latter form is closely reminiscent of elliptic equations extensively studied in relation to classic `prescribed scalar curvature' problems of differential geometry (though the exact power appearing in the power-law nonlinearity is different). 

If one aims at constructing a massless Klein-Gordon (i.e., wave) equation corresponding to the original quantum-mechanical system, the nonlinearity drops out, resulting in a much simpler problem. In this case, known ground state wavefunctions for the original quantum system can be utilized for the conversion procedure, as described in section \ref{massless}. We have demonstrated how this approach works for superintegrable Rosochatius systems (\ref{rosH}), resulting in a family of spacetimes (\ref{rosmetr_final}) resonant with respect to the massless wave equation.

We conclude with a list of open questions relevant for our formalism:
\begin{itemize}
	\setlength\itemsep{1em}
	\item General theory of existence of solutions of (\ref{yma}) would contribute appreciably to clarifying the operation of our formalism. Similar equations arising in differential geometry \cite{prescr} have been thoroughly analyzed, hence one should expect that the situation for our equation may as well be elucidated.
	\item In practical applications of our formalism, we have focused on the case of zero Klein-Gordon mass, where (\ref{yma}) greatly simplifies. Are there any general technics for solving this equation (rather than analyzing the existence of solutions) for non-zero masses (at least, for solvable potentials in the original quantum-mechanical system).
	\item Equation (\ref{yma}) may in principle admit multiple solutions, given that there is freedom in choosing boundary conditions, depending on which conformal transformation one allows. Singular conformal transformations may also be allowed (and they may push boundaries at finite distance off to infinity). This is in fact the case for the AdS construction starting from the Higgs oscillator. It would be good to quantify this freedom in choosing solutions of (\ref{yma}) and understand which prescriptions result in spacetimes interesting from a physical perspective.
	\item Systems with quadratic spectra exist in extentions of the class of Hamiltonians we have considered here, given by (\ref{Hamlt}). For example, it is possible to include effects of monopole fields without distorting the spectrum \cite{monopole}. Klein-Gordonization is likely to generalize to such systems, resulting in Klein-Gordon equations with background gauge fields.
	\item It would be interesting to understand how the spacetimes resulting from our construction, such as (\ref{rosmetr_final}), function in the context of dynamical theories of gravity. For instance, Anti-de Sitter spacetime solves Einstein's equations with a negative cosmological constant. More complicated spacetimes may require some matter fields to be supported as solutions.
	
	 In the context of dynamical theories, the resonant linear spectra of our spacetimes will guarantee that weakly nonlinear dynamics of their perturbations is highly sophisticated. (Nonlinear instability of AdS, which is precisely a manifestation of such phenomena, is a broad currently active research area.)
	\item What are the symmetry properties of spacetimes generated by ``Klein-Gordonization''? 
	
	How do they connect to the symmetries of the original quantum-mechanical problem (and in particular hidden symmetries)? 
	
	Again, for the case of the Higgs oscillator, this perspective has turned out to be fruitful, and it would be good to see how it works in more general cases.
\end{itemize}


%% file: Conclusion.tex
\cleardoublepage
\phantomsection
\section*{Conclusion}

\addcontentsline{toc}{section}{Conclusion}

\label{sec:conclusion}

Let us summarize our results.
In \chap{\ref{sec:gr_memory}} we discussed how an impulsive signal in a singular hypersurface effects null geodesics in Minkowski space. A new approach has been suggested which allows for full analysis of geodesic congruences compared to previous studies \cite{Barrabes_Hogan-book, Barrabes:2001vy} where approximations were assumed. The method can be applied to any space-time and any geodesic congruence. It is based on the physically justified assumption that, in continuous coordinates, the geodesic vector of a test particle is continuous across the hypersurface. Thus, to obtain the geodesic vector in the future one just needs to apply a coordinate transformation on the geodesic vector in the past.

Applying this technique on a parallel null congruence in flat space we obtained the initial conditions for the congruence to the future of the shell. We proved that the resulting congruence stays hypersurface orthogonal, as it was before crossing the shell. Furthermore, we provided arguments which generalize this result to any hypersurface orthogonal congruence in the past, meaning that any such congruence will give rise to a hypersurface orthogonal congruence to the future. An equivalent physical statement would be that an impulsive signal does not effect the rotation of a congruence if it didn't rotate before crossing the shell. 

As stated above, in continues coordinates the geodesic vector flow does not suffer a jump upon crossing the shell. But a discontinuity arises in the gradient of the geodesic vector flow, the B-tensor. We have shown, for the parallel geodesic flow, that this discontinuity is related to different components of the stress-energy tensor. In particular the jump in the expansion is determined by the energy density and currents on the shell while the jump in the shear is determined by the gravitational wave component together with the surface currents. It is clear from \eqref{eq:dtheta} and \eqref{eq:dC} that the results are independent of the choice of congruence in the case of BMS supertranslations. This change in the B-tensor after the passage of an outgoing gravitational wave leads to a covariant description of the gravitational memory effect - the B-memory effect.

\chap{\ref{sec:Integrability_of_MP_1}} and \chap{\ref{sec:Integrability_of_MP_2}} were devoted to the integrability problem of Hamilton-Jacobi equation in the near horizon geometry of Myers-Perry black hole in arbitrary (even or odd) $d$ dimensions. The fully isotropic case in arbitrary dimensions has been fully studied previously. The fully non-isotropic problem in odd dimensions has been shown to be integrable, although the explicit form of the first integrals were unknown. We were able to introduce a convenient common description of the geometry in odd and even dimensions and unify these two cases into a single problem. After having this unified description, we continued studying the separability of variables in fully non-isotropic NHEMP geometry to prove the integrability of even dimensions as well and to find the explicit forms of the first integrals.

It was  shown in \cite{non-equal-general} that integrals of motion of Hamilton-Jacobi equation in fully non-isotropic case can be expressed through inverse Vandermonde matrix in ellipsoidal coordinates. Solving this equation we found the hidden symmetries and expressed them through initial coordinates. This procedure was explained in lower (7, 9 and 11) dimensions and a general formula for higher dimensional first integrals was derived. Using our unified description of odd and even dimensions it is trivial to extend the results to the even dimensions, where the problem is integrable as well. We also found the second rank Killing tensors generating these symmetries. It is also interesting to have a transformation relating the first integrals of fully isotropic NHEMP to the first integrals of fully non-isotropic NHEMP. Taking all rotation parameters to be equal to each other in the first integrals of fully non-isotropic NHEMP just transforms all of them to the Hamiltonian of the spherical mechanics of fully isotropic NHEMP, so finding a transformation between the first integrals of these two systems is not a trivial task. We found such a transformation by taking small variations of the rotation parameters from their isotropic value and introducing combinations of first integrals which commute with the isotropic Hamiltonian.

After finalizing the discussion of the special cases, meaning the fully isotropic and fully non-isotropic cases, it is apparent that they are a part of a bigger picture, the most general case when rotation parameters are grouped in blocks of equal and non-equal values. Indeed, it turns out that when some of the rotation parameters are equal to each other and are different from the rest, the system becomes superintegrable. In short, the steps for obtaining the hidden symmetries is the following. We start from a system with $2N+1+\sigma$ variables with $N$ isometries. Fixing the momenta associated with the isometries, we obtain and focus the $N-1+\sigma$ dimensional ``angular mechanics'' part. By introducing a special coordinate system, which is a mixture of spherical and ellipsoidal coordinates, we separate the variables in the angular mechanics thus introducing $N-1+\sigma$ independent constants, or the first integrals.
This system reduced to its special cases of fully isotropic and fully non-isotropic NHEMP after appropriate assumptions.

The next step in our discussion was the extremal vanishing horizon geometry, which exists in odd dimensions, when one of the  rotation parameters $a_i$ vanishes. In the general NHEVHMP case, where the background isometry is $SO(2,2)\times U(1)^{\frac{d-3}{2}}$ the number of independent charges associated with Killing vectors is $\frac{d+1}{2}$. The system contains two conformal algebras, but they have the same Casimir operator, so there is a single angular mechanics. As a result the system remains integrable and no new independent constants of motion exist compared  to the non-vanishing case.

In \chap{\ref{sec:klein_gordon}} we have suggested an approach for mapping quantum systems to a Klein-Gordon equation on a curved space-time. In general, the procedure is the following. We start from the equation \eqref{KGwannabe} which defines a scalar field and can be reduced to the Schr\"odinger equation with Hamiltonian \eqref{Hamlt} after separating the time variable. Equation \eqref{KGwannabe} is not a Klein-Gordon equation yet but can be transformed into such after an appropriate conformal rescaling \eqref{eq:conf_resc} of the metric and the scalar field. Such a conformal factor should satisfy a non-linear elliptic equation \eqref{confgamma}, which greatly simplifies with a further assumption of the Klein-Gordon equation being massless. Now, we are primarily interested in systems with quadratic spectra for various reasons mentioned in the introduction and in the \chap{\ref{sec:klein_gordon}}. We have shown that for these spectra the ground state wavefunction of the initial Schr\"odinger equation satisfies the elliptic equation for the conformal factor. In other words, the ground state wavefunction of the initial Schr\"odinger equation defines the conformal factor which maps the quantum system to a massless wave equation.

Many well-known physical systems, including superintegrable ones, have quadratic spectra. Examples include the Higgs oscillator, the superintegrable Rosochatius system and elementary systems like one dimensional infinite rectangular potential well problem and the trigonometric P\"oschl-Teller system. We have demonstrated how the proposed mapping procedure can be applied on the Higgs oscillator and the superintegrable Rosochatius system. In the case of Higgs oscillator, the procedure results into a massive Klein-Gordon equation in the AdS background. In the case of superintegrable Rosochatius system we obtain a massless field equation on the background \eqref{rosmetr_final}.

\newpage

\cleardoublepage
\phantomsection
\section*{Summary}
\addcontentsline{toc}{section}{Summary}
Here we present the outline of the main results of this thesis.

\begin{itemize}
	\item A new approach has been suggested for studying the effects of impulsive gravitational waves of congruences encountering them.
	\item The technique has been applied on null congruences. It has been established that hypersurface orthogonal null congruences stay such after crossing the shell.
	\item A covariant definition of the gravitational memory effect has been suggested based on the B-tensor of the congruence. The relations between the components of the B-tensor and the stress-energy tensor of the shell have been derived.
	\item The B-tensor has been calculated and the approach has been demonstrated for BMS type soldering.
	\item A common description has been introduced for even and odd dimensional NHEMP geometries.
	This description was used to prove that the even dimensional fully non-isotropic NHEMP system is integrable. 
	\item Integrals of motion, as well as the Killing vectors of the fully non-isotropic NHEMP in arbitrary dimensions have been presented in initial coordinates.
	\item We found a non-trivial transformation between the integrals of motion of fully non-isotropic and fully isotropic NHEMP black hole geometries.
	\item We separated the variables of the most general partially isotropic NHEMP and showed its transformation to the special cases of fully non-isotropic and isotropic NHEMP.
	\item A new approach has been suggested for mapping Schr\"odinger equation on a curved background to a Klein-Gordon equation on the background of another geometry. 
	\item We have shown that this procedure greatly simplifies for systems with quadratic spectra and applied it on the Higgs oscillator and the superintegrable Rosochatius system.
	
\end{itemize}

%% file: Acknowledgement.tex
 \section*{Acknowledgements}
\label{sec:acknowledgements}

Firstly, I would like to express my sincere gratitude to my advisor Prof. Armen Nersessian for the continuous support of my Ph.D study and related research, for his patience, motivation, and immense knowledge. He encouraged me to grow as an independent scientist and his guidance helped me in all the time of research and writing of this thesis. Besides everything, he made it possible for me to visit many institutions abroad and cooperate with foreign scientists. He also made me part of many grants and projects, the financial support from which allowed me to fully concentrate on my research. 

My sincere thanks goes to Prof.  Martin O'Loughlin who first supervised my master's thesis and then, along with Prof. Nersessian, part of my Ph.D research. Despite his busy schedules, Prof. O'Loughlin has made himself available for scientific discussions and was always open to share with me his profound research experience.

I also thank Prof. M.M. Sheikh-Jabbari for giving me the opportunity to visit the Institute for Research in Fundamental Sciences (IPM) and work with him and his research team in a very friendly and motivating environment. I greatly appreciate the collaborative work with Prof. Oleg Evnin and thank him for his proposed problem to which a chapter of the current thesis is dedicated and for his guidance for solving it.

Last but not the least, I would like to thank Byurakan Astrophysical Observatory staff members and researchers for their continues support, patience and encouragement. They have assisted me with any problem and question and backed me from the very first day I entered the observatory.

%% file: Bibliography.tex
\cleardoublepage
\phantomsection